\documentclass[journal,draftcls,onecolumn,11pt,twoside]{IEEEtran}

\usepackage{xcolor}
\usepackage{multirow}

\usepackage{array}
\usepackage{soul}
\usepackage{cite}
\usepackage[utf8]{inputenc} 
\usepackage[T1]{fontenc}    
\usepackage{hyperref}       
\usepackage{url}            
\usepackage{booktabs}       
\usepackage{amsfonts}       
\usepackage{nicefrac}       
\usepackage{microtype}      
\usepackage{comment}
\usepackage{amsthm}
\usepackage{pgfplots}
\usepackage{amsmath}
\usepackage{algorithm}
\usepackage{algpseudocode}
\usepackage{amssymb}
\usepackage{xcolor}

\newtheorem{theorem}{Theorem}
\newtheorem{lemma}[theorem]{Lemma}
\newtheorem{definition}{Definition}
\newtheorem{remark}{Remark}
\usepackage{subcaption}
\usepackage{tikz}

\usepackage{amsfonts}
\usepackage{dsfont}
\usepackage{amsthm}

\usepackage{microtype}      
\usepackage{xcolor}         

\usepackage{float}
\usepackage{amsmath}
\usepackage{dsfont}

\usepackage {tikz}
\usetikzlibrary{matrix}
\usetikzlibrary {positioning}
\usetikzlibrary{calc}

\usepackage[mathscr]{eucal}
\usepackage{cancel}

\def\Figurespath{./images}

\title{Track-MDP: Reinforcement Learning for Target Tracking with Controlled Sensing}
\author{Adarsh M. Subramaniam,~\IEEEmembership{Graduate Student Member,~IEEE}, Argyrios Gerogiannis,~\IEEEmembership{Graduate Student Member,~IEEE}, James Z. Hare,~\IEEEmembership{Member, ~IEEE}, and  Venugopal V. Veeravalli,~\IEEEmembership{Fellow, ~IEEE}

\thanks{This work  was supported in part by the DEVCOM Army Research Laboratory under Cooperative Agreement W911NF-17-2-0196, through the University of Illinois at Urbana-Champaign.}

\thanks{Subramaniam, Gerogiannis and Veeravalli are with the ECE Department and the Coordinated Science
Lab, University of Illinois at Urbana-Champaign, Urbana, IL, USA. Hare is with DEVCOM Army Research Laboratory, Adelphi, MD, USA. Email: \{adarshm2, ag91, vvv\}@ILLINOIS.EDU, james.z.hare.civ@ARMY.MIL. 
}
}
\begin{document}
\maketitle
\begin{abstract}
State of the art methods for target tracking with sensor management~(or controlled sensing) are model-based and are obtained through solutions to Partially Observable Markov Decision Process~(POMDP) formulations. 
In this paper a Reinforcement Learning (RL) approach to the problem is explored for the setting where the motion model for the object/target to be tracked is unkown to the observer. It is assumed that the target dynamics are stationary in time, the state space and the observation space are discrete, and there is complete observability of the location of the target under certain (a priori unknown) sensor control actions. Then, a novel Markov Decision Process~(MDP) rather than POMDP formulation is proposed for the tracking problem with controlled sensing, which is termed as Track-MDP. In contrast to the POMDP formulation, the Track-MDP formulation is amenable to an RL based solution. It is shown that the optimal policy for the Track-MDP formulation, which is approximated through RL, is guaranteed to track all significant target paths with certainty.
The Track-MDP method is then compared with the optimal POMDP policy, and it is shown that the infinite horizon tracking reward of the optimal Track-MDP policy is the same as that of the optimal POMDP policy. In simulations it is demonstrated that Track-MDP based RL leads to a policy that can track the target with high accuracy. 
\end{abstract}
\section{Introduction}
We consider the problem of target tracking in a sensor network, where the sensing process can be controlled to optimize tracking performance under resource constraints. In particular, the target moves through a sensing region and its location is estimated based on the data collected from a set of active sensors. Turning on all sensors in the grid results in high tracking accuracy but dramatically low energy efficiency, and vice versa. An ideal control input selects a subset of sensors to be turned on based on past observations that optimizes {the tradeoff} between tracking accuracy and energy efficiency~\cite{pose_zach,target_loc}. {We refer to this problem as target tracking with controlled sensing (TTCS).}

Existing solutions to the TTCS problem typically assume that the target's motion model is known in advance. These solutions then select control inputs for the system to activate a subset of sensors at each time instance to enable the estimation of the target's position. This model based TTCS problem is a well-studied and has applications in intelligence, surveillance, and reconnaissance (ISR) operations \cite{pose_zach,jpda,tracking_sensor_deploy}. Examples of motion model assumptions include constant velocity, constant acceleration, and coordinated turn models \cite{BarShalom2001EstimationWA,BarShalom_handbook}. The tracking system uses the known motion model to estimate the state of the target using filtering algorithms, such as the Extended Kalman Filter~(EKF)~\cite{moulin_veeravalli_2018} or Unscented Kalman Filter~(UKF)~\cite{ukf}. The entire approach relies on the fact that the model is accurately known; an unrealistic assumption in adversarial settings. 

In this work, we study the TTCS problem without any prior knowledge of the motion model. To this end, we model the target's motion as a Markov process in a discrete state space, with a probability transition matrix that is \emph{unknown} to the tracking system.  At each time step, the target to be tracked transitions to a new position, and the position of the target is determined based on the measurements from the active sensors~(sensors that have been turned on). The sensors that are turned on are determined by a control input given by a central controller that attempts to optimize the tradeoff between resource utilization and tracking accuracy. 
In this setting, the observations from the active sensors may be noisy or uninformative resulting in partial observability of the target. Thus, TTCS is a {\emph{Partially Observable Markov Decision Process}}~(POMDP) problem. 

Solutions to the POMDP arising in the TTCS problem have been proposed in~\cite{fuemmeler_sleepcontrol,fuemmeler_smartsleep,hero2011sensor} under the assumption that the target's motion model is known. 
In the finite horizon setting with known target movement model, an  optimal POMDP policy for the TTCS problem can be found by decomposing the value function into hyperplanes corresponding to each action and choosing the action hyperplane that maximizes the value function \cite{ross2008online}.  For the case where target movement model is unknown, the framework of \cite{spectral_anima} can be applied, in which estimates of the transition and observation kernels are obtained through spectral
methods and these estimates are used to compute a policy. However, it is to be noted that this policy is guaranteed to have a sub-linear regret only with respect to the optimal
memoryless POMDP policy and not the optimal POMDP policy.  The TTCS POMDP problem with unknown target movement model is studied in \cite{particlefilter_pomdp_edwin}, where a particle filter is used to estimate the target transition kernel. We emphasize that policies based on solving the POMDP arising in the TTCS problem are computationally intensive even with knowledge of the target movement model \cite{papa_tsitsiklis}, and the solutions become even more intractable when the model is unknown.



In this work, we address the challenge of finding the optimal POMDP policy for TTCS, by proposing a surrogate MDP framework which we term the \emph{Track-MDP} framework. This framework leverages MDPs, thus offering advantages such as polynomial-time computation~\cite{littman_complexity} of the optimal policy when the model is known, and sublinear regret~\cite{ucrl} when the model is unknown. We demonstrate that the Track-MDP's optimal policy achieves accurate target tracking and establish that its infinite horizon tracking reward is the same as that of the optimal POMDP policy. Given the computational intractability of computing the optimal POMDP policy, we compare the Track-MDP's optimal policy, approximated via RL, with the popular $\mathrm{Q}_{\mathrm{MDP}}$ policy~\cite{littman_qmdp} in simulations. 

%
\subsection{Contributions}
\begin{enumerate}
    \item 
    We introduce the Track-MDP framework, which is the first MDP framework for the TTCS problem. We further define the reward function and transition dynamics of the Track-MDP, and study the properties of the optimal RL policy on the MDP.
    \item We establish the \emph{track property} which asserts that the optimal policy of the Track-MDP is guaranteed to track every potential path of the target that has a probably exceeding a specified threshold. This threshold is {dependent} on the reward parameters of the Track-MDP.
    \item We show that the infinite horizon reward of the optimal policies of the Track-MDP and the POMDP are equal for the {TTCS} problem. 
    \item In our experiments we employ RL algorithms to approximate for the optimal policy on the Track-MDP. We show that the obtained Track-MDP policy can outperform the $\mathrm{Q}_{\mathrm{MDP}}$ policy~\cite{littman_qmdp} for the POMDP in terms of tracking accuracy as well as in the infinite horizon tracking reward. 
\end{enumerate}
The remainder of the paper is as follows. In Section~\ref{sec:problem_form}, we formulate the TTCS problem. In Section~\ref{sec:pomdp_formulation}, we review the  POMDP literature addressing the TTCS problem. In Sections~\ref{sec:tmdp_for}- \ref{sec:equivalence}, we introduce our RL solution. In Section~\ref{sec:theory}, we analyze the performance RL on the Track-MDP, and compare it with the optimal POMDP policy. In Section~\ref{sec:exp}, we provide some experimental results to demonstrate the effectiveness of our approach.
\subsection{Notation} 
 {We use $|C|$ to denote the cardinality of a set $C$, $\mathds{1}_{\{B\}}$ to denote the indicator function of an event $B$, and $2^A$ to denote the powerset of a set $A$. Additionally, we use $e_{b^j}$ to denote the unit vector of certain length with zeros in every position apart from the $j-$th position, and $[c]_j$ to denote the $j-$th element of the vector $c$.  We use $[M]$ to denote the set $\{1,2,\cdots,M\}$, and represent by $\Delta(n)$, the $n-1$ dimensional probability simplex consisting of valid probability vectors of length $n$. }
\section{Problem Formulation}\label{sec:problem_form}
Consider the discrete-time dynamical system, with hidden state $x_k$, where $k\in \mathbb{N}$ denotes the time step. The hidden state of the system evolves as
\begin{equation}\label{eq:process}
\begin{split}
x_{k+1}=f(x_k,w_k),\\
\end{split}
\end{equation}
where $f$ is an unknown state-transition function and   $w_k$ is the process noise. The observation $y_k$ is a function $g$ of the control action $a_{k}$, hidden state $x_k$ and a measurement noise $\mu_k$. Mathematically,  
\begin{equation}\label{eq:observation}
\begin{split}
y_{k}=g(x_k,a_{k},\mu_k).\\
\end{split}
\end{equation}
 In controlled sensing, the objective is to choose a sequence of actions $\{a_i\}_{i=1}^{k}$ and estimate the state $x_k$ as a function of the sequence $\{y_i\}_{i=1}^{k}$ of observations.
\subsection{Simplified Controlled Sensing Problem}
In this section, we simplify the controlled sensing problem to function as a mathematical framework for tracking the hidden state $x_k$ via sensor management on a discretized sensing region outlined in Section~\ref{sec:sensing_grid}. The simplified process model and observation model are defined as follows.
\subsubsection{Simplified Process Model}~\label{sec:simplified_process_model}
The evolution of the hidden state $x_k$ in~\eqref{eq:process} is modeled by a finite state space discrete time Markov chain ${X}_k$ with state space $\mathcal{X}$, defined as 
\begin{align}\label{eq:process_states}
    \mathcal{X} \triangleq \mathcal{\bar{X}} \cup \{ b^T\},\ 
    \mathcal{\bar{X}} \triangleq \{b^i : i \in [N(S)] \}.
\end{align}
%
%
In~\eqref{eq:process_states}, the elements of the state space $\mathcal{X}$ are labelled as $b^i$ for $i\in [N(S)]$, where $N(S)$ is defined as the cardinality of $\bar{\mathcal{X}}$
and  $b^T$ is termed the terminal state. Hence, the cardinality of the state space $|\mathcal{X}|$ is $N(S)+1$. The evolution of the hidden state $x_k$ is described by a stationary transition probability matrix $P$ of dimension $(N(S) +1) \times (N(S)+1)$, where 
\begin{equation} \label{eq:process_st_transition}
 P_{i,j} = \begin{cases}
			\mathrm{Pr}(x_{k+1} = b^j | x_k = b^i ), & \text{if $i,j \in [N(S)]$ }\\[5pt]
            \mathrm{Pr}(x_{k+1} = b^T | x_k = b^i ), & \vtop{\hbox{$\mbox{if}\:\:j = N(S) +1,$}\hbox{$i \in [N(S)]$}}\\[15pt]
             \mathrm{Pr}(x_{k+1} = b^j | x_k = b^T ), & \vtop{\hbox{$\mbox{if}\:\:i = N(S) +1,$}\hbox{$j \in [N(S)]$}}\\[15pt]
              \mathrm{Pr}(x_{k+1} = b^T | x_k = b^T ), & \vtop{\hbox{$\mbox{if}\:\:i = N(S) +1,$}\hbox{$j = N(S) +1$}}
\end{cases}.
\end{equation}
Then the terminal state $b^T$ has the property that $$ \mathrm{Pr}(x_{k+1} = b^T | x_k = b^T ) = 1.$$
 \subsubsection{Simplified Observation Model}
Consider a finite discrete set $\mathcal{A}_{\mathcal{\bar{X}}}$ of control actions defined as 
\begin{align}\label{eq:measurment_actions}
    \mathcal{A}_{\mathcal{\bar{X}}} \triangleq \{a : a \in 2^{\mathcal{\bar{X}}} \},
\end{align}
 The control action $a_k$ is an element of the set $\mathcal{A}_{\mathcal{\bar{X}}}$, and the observation $y_k$ in~\eqref{eq:observation} is given by
\begin{equation}\label{eq:measurement_observations}
 y_k= \begin{cases}
     x_k & \text{if $x_k \in a_{k}$}\\
     {\mathcal{E}} &\text{if $x_k \notin a_{k}$}.
 \end{cases}  
\end{equation}
The hidden state $x_k$ is observed without any noise if $x_k \in a_{k}$ and the observation is completely noisy or uninformative if $x_k \notin a_{k}$. We represent the uninformative observation by $\mathcal{E}$. 

From the definition of the state space $\mathcal{X}$, the observations $y_k$ belong to the set $\mathcal{Y}$, where 
\begin{align}\label{eq:observation_space}
    \mathcal{Y}= \{b^j:j \in [N(S)]\} \cup \{ \mathcal{E} \}.
\end{align}
\subsection{Sensing Grid}~\label{sec:sensing_grid}
\begin{figure}[H]
        \centering
\resizebox{150 pt }{150 pt}{
\begin{tikzpicture}[line width=.7pt, outer sep=0pt]

\draw[step=1.2cm,black] (0.0,0.0) grid (6,6);
\draw [red,thick] plot [smooth,tension=1] coordinates { 
   (0,0) (1.5,3) (4.5,3) (5,5) (6,6)};
\newcounter{mycounter};
\setcounter{mycounter}{0}; 
\foreach \y in {1,...,5}
  \foreach \x in {1,...,5}
    \addtocounter{mycounter}{+1} 
    \draw [fill=blue](\x*1.2-0.6,\y*1.2-0.6) circle (0.1cm) node[above left ] {$b^{\themycounter}$} ;

\end{tikzpicture}
}
\caption{Representation of tracking in a sensing grid.}
        \label{code1}
\end{figure}
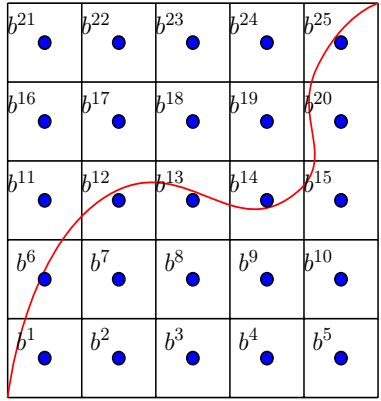

 {We consider a rectangular sensing region, which is discretized into $N^2$ units termed cells, as illustrated in Figure~\ref{code1}. } {We model the motion of the target  with the simplified process model~\ref{sec:simplified_process_model}, by indexing the cells in the grid with the elements of the set $\bar{\mathcal{X}}$ with $N(S)=N^2$, i.e., the cells are denoted by labels $\{b^i\}_{i=1}^{N^2}$.}
%
%
Each grid cell $b^i$ has a sensor that outputs 1 if the target state $x_k$ is $b^i$, otherwise 0. A central controller selects action $a_k \in \mathcal{A}_{\Bar{X}}$ at time $k$ which is a subset of sensors in the grid positions $\{b^i\}_{i=1}^{N^2}$ to activate. From~\eqref{eq:measurement_observations}, if $x_k = b^i$ and $b^i$ is in $a_k$, the corresponding sensor at $b^i$ outputs 1, suggesting $y_k = x_k$. If $b^i \notin a_k$,  all sensor outputs are 0, implying $y_k=\mathcal{E}$. Additionally, a low-cost sensor detects target's entry and exit~($x_k = b^T$) and signals it unambiguously.

This formulation of the target motion and sensing model place the energy-efficient tracking problem within the controlled sensing framework, outlined by~\eqref{eq:process_st_transition} and \eqref{eq:measurement_observations}. Actions $\{a_k\}_{k\geq 1}$ are to be chosen to locate the target precisely while conserving energy. We formulate this through an infinite horizon reward.
\subsection{Tracking Reward on the Sensing Grid}
In a sensing grid, any tracking algorithm with sensor management must estimate the target's location $\hat{x}_k$ at each time step $k$. Typically, the mean squared error $||\hat{x}_k - x_k ||^2$ or Hamming error $\mathds{1}_{\{\hat{x}_k \neq  x_k\}}$  is considered as estimation error and must be minimized {(equivalently, the negative of the error should be maximized)}. Since the target's exact position $x_k$ is unknown, as in prior literature~\cite{fuemmeler_sleepcontrol,fuemmeler_smartsleep}, we aim to maximize a pseudo-reward
\begin{align}\label{eq:reward_func}
    R(x_k,a_{k}) &=   \big(r\cdot\mathds{1}_{\{x_{k} \in a_k \}}-c\cdot|a_{k}|\mathds{1}_{\{a_k \neq \hat{a}_s\}}  -D\cdot\mathds{1}_{\{a_k = \hat{a}_s\}}\big)
     \times \mathds{1}_{\{x_{k}\neq b^T\}},
\end{align}
where $r,c$ and $D$ are positive constants. The pseudo-reward, defined in~\eqref{eq:reward_func}, encourages target observation ($r\cdot\mathds{1}{\{x_{k} \in a_k }\}$), discourages excessive energy use ($c|a_{k}|$), and promotes safe sensing~(defined in Section~\ref{sec:assumption_constraint}), through the term $D\cdot\mathds{1}_{\{a_k = \hat{a}_s\}} $. The term $D\cdot\mathds{1}_{\{a_k = \hat{a}_s\}} $ adds a large negative reward $-D$ for turning on a {safe} sensing action $\hat{a}_s$.

Let the reward $R(x_{k},a_k)$ at time $k$ be denoted as $r_k$. We aim to maximize the Infinite Horizon Tracking Reward ($\mathrm{IHTR}$) defined in~\eqref{eq:inf_hori_rew}, representing the expected sum of rewards over infinite time steps, by choosing actions $a_k$ appropriately, given initial information $I_0$.
\begin{align}\label{eq:inf_hori_rew}
        \mathrm{IHTR}(I_0) \triangleq \mathbb{E}\left[\sum_{k=1}^{\infty} r_k\biggr\rvert I_{0}\right]= \mathbb{E}\left[\sum_{k=1}^{\infty} R(x_k,a_k)\biggr\rvert I_{0}\right].
\end{align}
\subsection{Constraints and Assumptions}\label{sec:assumption_constraint}
In this section we describe the constraints on tracking algorithms deployed on the sensing grid~\ref{sec:sensing_grid} and assumptions on the initial information $I_0$ provided by the sensing grid to tracking algorithms.   

\noindent \textbf{$T_{\max}$ safe sensing constraint:} 
A tracking algorithm on the sensing grid is labeled $T_{\max}$ safe if, within any time interval of length $T_{\max} + 2$, the target is perfectly located ($y_k \neq \mathcal{E}$) at least once. If the target remains untracked ($y_k = \mathcal{E}$) for $T_{\max} + 1$ time steps, a designated sensing action $\hat{a}_s$ is triggered to observe the target with certainty. Examples of $\hat{a}_s$ include activating a high-power sensor or all sensors in the grid. This safe sensing condition ensures consistent monitoring of tracked targets, which is crucial especially for adversarial scenarios requiring frequent observation.

\noindent\textbf{Initialization of Tracking Algorithms on the Sensing Grid:}\label{sec:sensing_grid_init}
In this study, the sensing grid includes a low-power sensor {for} detecting target presence and a high-power sensor or action for accurate target localization. Upon target entry, the low-power sensor detects its presence, followed by activation of the high-power sensor to determine its precise location, denoted as $x_0$. This marks time instance $k=0$, with $x_0$ serving as the initial position input, or initial information ($I_0$), for any deployed tracking algorithm on the grid.
\subsection{Target Tracking on Sensing Grid: An Example}
Consider the scenario where the target is moving in the $5\times 5$ sensing grid shown in Figure~\ref{code1}. The state space $\mathcal{X}$ is $\mathcal{X} = \{b^i | i\in \{1,2,\cdots 25\}\} \cup \{b^T\}.$
At time $k=1$, let the state/position of the target be $x_1 = b^{13}$. If $a_1 =\{b^{13}, b^{5}, b^{4} \} $, then the observation $y_1 = b^{13}$ since $b^{13} \in a_1$. However, if $a_1 = \{b^{5}, b^{4} \}$, the observation $y_1 = \mathcal{E}$~(completely uninformative) since $b^{13} \notin a_1$. The position of the target at time $k=2$ denoted as $x_2$ evolves according to some stationary Markov chain $P$ as defined in~\eqref{eq:process_st_transition}. Additionally, if at time $k'$, the target exits the grid, then $x_{k'} = b^{T}$. A low powered sensor detects this exit of the target from the grid without any ambiguity.

With the problem formulated and the objective set as maximizing the IHTR, we introduce {some additional} notation. Then, we explore the POMDP formulation of the energy-efficient tracking problem in literature.
\subsection{{Additional} Notation}
 Let ${\alpha}$ be a probability distribution over the elements of the set $\bar{\mathcal{X}}$ (i.e. ${\alpha} \in \Delta(N^2)$). Since the elements $\{b^j\}_{j=1}^{N^2}$ of the set $\bar{\mathcal{X}}$ correspond to the cells on the sensing grid, $\alpha$ is a probability distribution over the cells of the sensing grid. Let $ a \in \mathcal{A}_{\Bar{\mathcal{X}}}$ be a sensing action. For the action $a$, define the set $\mathcal{B}_a = \{j: b^j \in a\}$.

Define $\tilde{\alpha}\triangleq [\alpha]_{\Bar{a}}$. The vector $\tilde{\alpha}$ is also a probability distribution over the cells of the sensing grid and is defined as
\begin{equation*}
 [\tilde{\alpha}]_j= \begin{cases}
     \frac{[\alpha]_j}{(1-\sum_{j\in B_a}[\alpha]_j)} & \text{if $j \in [N^2]\setminus B_a $}\\
     0 &\text{if $j \in B_a$}
 \end{cases}  .
\end{equation*}
\textit{Example:} Consider a $N\times N$ grid with $N=2$. The grid cells are labelled $\{b^1,b^2,b^3,b^4\}$. Let $\alpha = (0.2,0.3,0.4,0.1) $ and $a=\{b^1\}$. Then, $\tilde{\alpha}= [\alpha]_{\Bar{a}}$ is 
\begin{align}
    [\alpha]_{\Bar{a}} = \left(0,\frac{0.3}{(1-0.2)},\frac{0.4}{(1-0.2)},\frac{0.1}{(1-0.2)}\right).
\end{align}
 The vector $e_{b^j}$ is a probability distribution over the cells $\{b^j\}_{j=1}^{N^2}$ of the sensing grid such that $[e_{b^j}]_j = 1$ and $[e_{b^j}]_i = 0$ for $i \neq j$. Denote by $\tilde{a}$, the action of turning on all sensors in the grid and $\tilde{a}_{b^j}$ be the action of turning on all sensors in the grid except sensor at $b^j$. With a slight abuse of notation, $\pi$ is employed to represent the policy for both a POMDP and the Track-MDP. It is inferred from the context, whether the domain of $\pi$ is the Track-MDP state $s_k$ or the POMDP sufficient statistic $v_k$. 
 
%
\section{The POMDP Formulation}~\label{sec:pomdp_formulation}
The TTCS problem falls under the POMDP framework, since the location of the target $x_k$ is unknown when the observation $y_k$ is uninformative~(i.e. $y_k = \mathcal{E}$)~\cite{fuemmeler_sleepcontrol}. 

For the POMDP formulation of the tracking problem, the process state space is the set $\mathcal{X}$ with $N(S)= N^2$ and the observation space is the set $\mathcal{Y}$ as defined in Section~\ref{sec:problem_form}.

In the POMDP formulation for the {TTCS problem } studied in~\cite{hero2011sensor,fuemmeler_sleepcontrol,fuemmeler_smartsleep}, the objective is to maximize the $\mathrm{IHTR}$ by choosing suitable actions $a_k \in \mathcal{A}_{\mathcal{\bar{X}}}$ at each time $k$. The sensing action at time $k+1$ is allowed to be a function $\pi_k$ of the information available at time $k$. This information, defined as $I_k$, is the ensemble of all the past observations and actions. Mathematically,
\begin{align}~\label{eq:info_increase}
    &I_k \triangleq \{y_0,\cdots,y_k,a_1,\cdots,a_{k}\} , \ a_{k+1} = \pi_k(I_k).
\end{align}
 Executing the action $a_{k+1}$ turns on sensors in the sensing grid and depending on the position of the target $x_{k+1}$, an observation $y_{k+1}$ is made according to~\eqref{eq:measurement_observations}. Subsequently, the information is updated as $I_{k+1} = \{y_1,\cdots,y_k,y_{k+1},a_1,\cdots,a_{k},a_{k+1}\}$.
The next control input $a_{k+2}$ for sensing the target at time $k+2$ is determined by $\pi_{k+1}(I_{k+1})$ and so on. The sequence of control inputs $\pi_0,\pi_1,\cdots,$ is defined as the policy $\pi$ of the POMDP formulation. We represent the policy $\pi$ as an infinite dimensional vector with $[\pi]_i = \pi_i$. As shown in~\eqref{eq:info_increase}, the information $I_k$ expands with time, requiring potentially infinite memory storage. {It is easy to show (see, e.g., \cite{sondik_pomdp}) that the probability distribution of state $x_k$ given $I_k$, termed the belief state $\beta_k$, is a sufficient statistic for the POMDP.} On the tracking grid studied here,
\begin{align}
    [\beta_k]_j = P(x_k = b^j | I_k),\ 
    [\beta_k]_{N^2 +1} = P(x_k = b^T | I_k).
\end{align}
Note that the action $a_k$ is set to $\hat{a}_s$ if the target has been unobserved for $T_{\max}+1 $ time steps. As a result, the sufficient statistic/belief at time $k$ must include information about the time since the last observation, denoted by $n(k)$. {We therefore append $\beta_k$ with $n(k)$ to form the tuple $v_k \triangleq (\beta_k, n(k))$, which we will henceforth refer to as the belief, with some abuse of terminology}. Mathematically, the term $n(k)$ can be defined as 
\begin{align}~\label{eq:lk_nk}
    l(k) & \triangleq \max \{p : \ 0\leq p \leq k,\  p\in \mathbb{N}, \ y_p \neq \mathcal{E} \} ,&&\\
    \nonumber n(k) &\triangleq k-l(k). 
\end{align}
If a sensing action $a_{k+1} = \pi_k(v_k)$ is executed at time $k+1$ to sense the target and if the observation model is as defined in~\eqref{eq:measurement_observations}, the belief $v_{k+1} = (\beta_{k+1},n(k+1))$ evolves as 
\begin{flalign}\label{eq:belief_up}
    &\beta_{k+1} = \Big(\big(\mathds{1}_{\{y_{k+1} = \mathcal{E} \}} [\beta_k P]_{\overline{a_k}} +  \mathds{1}_{\{y_{k+1} = x_{k+1}\}} e_{x_{k+1}}\big) &&\\
     &\nonumber\hspace{3em}\times  \mathds{1}_{\{n(k)\neq T_{\max}+1\}}+ \mathds{1}_{\{n(k)= T_{\max}+1\}} e_{x_{k+1}}\Big)\mathds{1}_{\{x_{k+1}\neq b^T\}}&&\\
     &\nonumber\hspace{3em}+\mathds{1}_{\{x_{k+1} = b^T\}} e_{N^2+1},&&\\
&\nonumber n(k+1) = \mathds{1}_{\{n(k)\neq T_{\max}+1\}}\mathds{1}_{\{y_{k+1} = \mathcal{E} \}} (n(k)+1).
\end{flalign}
Subsequently, the sensing action at time $k+2$ is $a_{k+2} = \pi_{k+1}(v_{k+1})$. Having defined the evolution of the belief, the IHTR for the policy $\pi$ is 
\begin{align}\label{eq:pomdp_value}
    V^{\pi,P}_{\mathrm{POMDP}}(v_0) \triangleq \mathbb{E}\left[\sum_{k=1}^{\infty} R(x_{k},a_{k})\biggr\rvert v_0\right].
\end{align}
It is well know in POMDP literature that when the underlying Markov process $X_k$ is  stationary, the policy that maximizes~\eqref{eq:pomdp_value} is also stationary~(i.e. $\pi_0 = \pi_1 = \cdots = \pi^*$) and is a function of the sufficient statistic $v_k$. We denote this optimal policy as $\pi^{*}_{\mathrm{POMDP}}$. Note that
\begin{align}
    a_{k+1}^* = \pi^{*}_{\mathrm{POMDP}}(v_{k}).
\end{align}
We discuss next the Track-MDP formulation, which is the main contribution of this work.  
\section{Track-MDP Formulation}~\label{sec:tmdp_for}
%
%
%
{The POMDP formulation for the TTCS problem requires knowledge of the transition probability matrix $P$ to update $v_k$. To learn to track the target without knowledge of $P$ via RL, we propose an MDP framework, which we term Track-MDP.} Below, we define Track-MDP rigorously for the sensing grid outlined in Section~\ref{sec:sensing_grid}.

\subsubsection{Action Space ($\mathcal{A}$)} 
The action space of the Track-MDP  is the {is the same as that of the POMDP, i.e., the} set $ \mathcal{A}_{\mathcal{\bar{X}}}$ with 
$\mathcal{\bar{X}} = \{b^i\}_{i=1}^{N^2}$.
\subsubsection{State Space ($\mathcal{S}$)}
A state of the Track-MDP is a tuple comprising of three elements.  
The state space consisting of all the possible states is the set,
\begin{equation*}
\begin{split}
\mathcal{S}=\big\{ (x,n,h) | x\in \mathcal{X},\: 0\leq n \leq T_{\max}+1,\: h\in \mathcal{A}^n \big\}.\\ 
\end{split}
\end{equation*}
The state $s_k$ at time $k$ is defined as
\begin{equation}\label{eq:state_des}
    s_k=(x_{l(k)},n(k),h_k) ,   
\end{equation}
where, $h(k) = \{ a_{l(k)+1},\cdots, a_{k} \}$ and $l(k),n(k)$ are as defined in~\eqref{eq:lk_nk}. In the expression of the state, $l(k)$ is the last time step at which the state was perfectly observed, $n(k)$ is the time since the last perfect observation and $h_k$ is the history of actions since the last perfect observation. The cardinality of the state space $|\mathcal{S}|$ is finite and is parametrized by $T_{\max}$.

\textit{Example to illustrate state space:} Let the target to be tracked on the sensing grid at time $k=4$ be at position $x_4 = b^{13}$. Let $a_4 = \{b^{13},b^{18},b^{20}\}$. Since $x_4 \in a_4$, the observation $y_4 = x_4$. From the description of the state in~\eqref{eq:state_des}, $s_4 = (x_4,0,\emptyset)$. Let it be the case that the sensing actions $a_5,a_6$ are such that $ x_5 \notin a_5$ and $x_6 \notin a_6$. At time $k= 6$, the state of the Track-MDP is $s_6 = (x_4,2,\{a_5,a_6\})$.    
\subsubsection{Transition Dynamics ($\mathcal{P}$)}
The state transition probability $\mathcal{P}(s_{k+1}|s_k,a_{k+1})$ implicitly depends on the target transition matrix $P$. To elucidate this implicit dependence, we describe the temporal evolution of the state $s_k$.

The state of the Track-MDP at time $k+1$, denoted by $s_{k+1}$ is a function of the current state $s_{k}=(x_{l(k)},n(k),h_k)$, action $a_{k+1}$ and observation $y_{k+1}$. It is given by 
\begin{equation} \label{eq:process_transition}
 s_{k+1} = \begin{cases}
			(x_{k+1},0,\emptyset), &\vtop{\hbox{$\mbox{if}\:\: y_{k+1}=x_{k+1} \:\: \mbox{or}$}\hbox{$n(k)=T_{\max}+1$}} \\[15pt]
   
            (x_{l(k)},n(k)+1,\{h_k,a_{k+1}\}), & \vtop{\hbox{$\mbox{if}\:\: y_k=\mathcal{E} \: \: \mbox{and} $}\hbox{$n(k)\leq T_{\max}$}}\\[15pt]
            
            s_T, & \mbox{if}\:\: x_{k+1}=b^T
\end{cases}.
\end{equation}
To further elaborate, for a given action $a_{k+1}$, if the observation $y_{k+1} = x_{k+1}$, the Track-MDP state transitions to $s_{k+1}=(x_{k+1},0,\emptyset)$. However, if the observation $y_{k+1} = \mathcal{E}$, the state transitions to $s_{k+1}=(x_{l(k)},n(k)+1,\{h_k,a_{k+1}\})$. If the target exits the grid, the target's position $x_{k+1}$ is $b^T$, leading to a transition of the Track-MDP state  from $s_k$ to $s_{k+1} = s_T$. If at time $k$, there was no observation of the target for $T_{\max}+1$ consecutive time steps in the past, the Track-MDP constrains the action to $\hat{a}_s$ due to safe sensing constraint and thus results in the state state transitioning to $s_{k+1}=(x_{k+1},0,\emptyset)$. 

From~\eqref{eq:process_transition}, it is evident that the state $s_{k+1}$ of the Track-MDP is a function of $s_k$, $a_{k+1}$ and $y_{k+1}$. For a given target transition matrix $P$, the distribution of $y_{k+1}$ is only a function of $s_k$ and $a_{k+1}$. This is true since the motion of the target is Markovian and $s_k = (x_{l(k)},n(k),h_k)$ corresponds to a probability distribution of the position of the target at time $k$, given that the target was at $x_{l(k)}$, $n(k)$ time steps back and the actions in $h_k$ not detecting the target. For example, $s_k = (b^{1},1,a)$ corresponds to the belief $[e_{b^{1}}P]_{\overline{a}}$.

From this argument, we infer that the distribution of $s_{k+1}$, is only a function of $s_k$ and $a_{k+1}$ and has the conditional probability distribution which we denote by $\mathcal{P}(s_{k+1}|s_k,a_{k+1})$. 
\begin{remark}
    The probability $\mathcal{P}(s_{k+1}|s_k,a_{k+1})$ is a function of the target transition kernel $P$. However, we omit the dependence on $P$ for notational convenience.  
\end{remark}
\subsubsection{Reward Function ($R$)}
The reward function of the Track-MDP at time $k$ is a function of states $s_{k-1}= (x_{l(k-1)},n(k-1),h_{k-1})$, $s_{k}= (x_{l(k)},n(k),h_{k})$ and the action $a_{k}$. The reward is defined as 
\begin{align}
    R(s_{k-1},a_{k},s_{k}) &= (r\cdot \mathds{1}_{\{ n(k)= 0\}} -  c|a_{k}|\mathds{1}_{\{ a_k \neq  \hat{a}_s\}}&&\\
    &\nonumber\hspace{1em} -D \cdot\mathds{1}_{\{ a_k =  \hat{a}_s\}}) \times \mathds{1}_{\{ x_k \neq  b^{T}\}}.
\end{align}
From the transition dynamics defined in~\eqref{eq:process_transition}, the term $n(k) =0$ is equivalent to the action $a_k$ sensing $x_k$~(i.e. $x_k\in a_k$). Hence, 
\begin{align}~\label{eq:reward_equi}
    R(s_{k-1},a_k,s_k) = R(x_k,a_k)
\end{align}
\textit{Example to illustrate reward function:} Let $s_{k-1} = (b^5,1,\{b^5,b^7\})$ and $a_{k} = \{b^9,b^{11}\}$. If $x_{k} = b^9$, then  $y_{k} = b^9$ and as a consequence $n(k)=0$. This implies that the reward is $R(s_{k-1},a_k,s_k) = r - c\cdot 2$. Alternatively, if  $x_{k+1} = b^7$, then $y_{k+1} = \mathcal{E}$ and as a consequence $n(k)= 2$. This implies that the reward $R(s_{k-1},a_k,s_k) =  -c\cdot 2$. 

\section{Tracking with a Policy on the Track-MDP}~\label{sec:tracking_with_tmdp}
Consider a policy $\pi:\mathcal{S} \rightarrow \mathcal{A}$ on the Track-MDP. Tracking the target with the policy $\pi$ works as follows. At time $k=0$, the sensing grid provides the policy with the information $I_0= x_0$ as detailed in Section~\ref{sec:assumption_constraint}. Correspondingly, the state of the Track-MDP is $s_0 =(x_0,0,\emptyset)$. The policy $\pi$ takes a sensing action $a_1 = \pi(s_0)$. At time $k=1$, the target moves to $x_1$ and the state of the Track-MDP evolves to either $s_1 = (x_1,0,\emptyset)$ or $s_1 = (x_0,1,\{a_1\})$, based on whether $x_1 \in a_1$ or $x_1 \notin a_1$. Depending on the observation, a reward $R(x_0,a_1)$ is obtained. The policy $\pi$ executes a sensing action $a_2 = \pi(s_1)$, and based on whether $x_2\in a_2$ or $x_2 \notin a_2$, the state of the Track-MDP further evolves at $k=2$ and rewards are obtained.
From~\eqref{eq:reward_equi}, the $\mathrm{IHTR}$ for a policy $\pi$ of the Track-MDP, defined as $V^{\pi,P}_{\mathrm{TMDP}}(s_0) $ is the same as~\eqref{eq:inf_hori_rew} with $I_0$ being the initial state $s_0$. 
%
%
%
Since the Track-MDP is an MDP formulation, there exists a policy $\pi$ that maximizes $V^{\pi,P}_{\mathrm{TMDP}}(s_0)$ for all possible $s_0$~\cite{sutton2018reinforcement}. We define this optimal policy as $\pi^*_{\mathrm{TMDP}}$ and an RL algorithm, such as Q-learning~\cite{sutton2018reinforcement}, Actor-critic~\cite{a2c}, Proximal Policy Optimization~\cite{ppo},  on the Track-MDP which is guaranteed to converge to $\pi^*_{\mathrm{TMDP}}$.
%
\section{Equivalence of Track-MDP States and POMDP Beliefs}~\label{sec:equivalence}
In this section, we establish a correspondence between Track-MDP states and POMDP beliefs, thus demonstrating an equivalence. We define a function $M_{\pi}:\mathcal{S}_{\pi} \rightarrow \Delta(N^2)$, where $ \mathcal{S}_{\pi}$ is a policy specific subset of the Track-MDP state space $\mathcal{S}$ and $\Delta(N^2)$ represents the $N^2-1$ dimensional probability simplex, within which probability distributions over the grid positions are defined.
 Through $M_{\pi}$, we establish an equivalence between Track-MDP states and POMDP beliefs while tracking an target. As a consequence, we prove that the $\mathrm{IHTR}$ of the optimal Track-MDP policy $\pi^*_{\mathrm{TMDP}}$ aligns with the $\mathrm{IHTR}$ of the optimal POMDP policy $\pi^*_{\mathrm{POMDP}}$.

\subsection{Definition of Function $M_{\pi}$}
The function $M_{\pi}$ is defined inductively as follows. If $n=0$, 
\begin{align}\label{eq:mpi_1}
        M_{\pi}((b^{k_0},n,\emptyset)) = (e_{b^{k_0}},0) \ \ \ \forall \ k_0 \in [N^2].
\end{align}
If $M_{\pi}((b^{k_0},n,h)) =  (\beta_n,n) \ \text{for some} \ 0\leq n\leq T_{\mathrm{max}}, \text{and}\ k_0 \in [N^2]$, then 
\begin{align}\label{eq:mpi_3}
     M_{\pi}((b^{k_0},n+1,\{h, \pi(v_n)\})) =  ([\beta_nP]_{\overline{\pi(v_n)}},n+1).
\end{align}
From the inductive definition of function $M_{\pi}$ in~\eqref{eq:mpi_1} and \eqref{eq:mpi_3}, it is evident that $M_{\pi}$ is defined for a subset of the Track-MDP states. We define this subset of states by $\mathcal{S}_{\pi}$. 
\subsection{Induced Policy on Track-MDP}\label{sec:induced_policy}
The function $M_{\pi}$ induces a policy on the states $\mathcal{S}_{\pi}$ of the Track-MDP. The induced policy $\tilde{\pi}$ is defined as
\begin{align}~\label{eq:induced_policy}
    \tilde{\pi}(s) = \pi(M_{\pi}(s))\ \forall s \in \mathcal{S}_{\pi}.
\end{align}
\begin{figure*}
\centering
\includegraphics[width=0.75\textwidth]{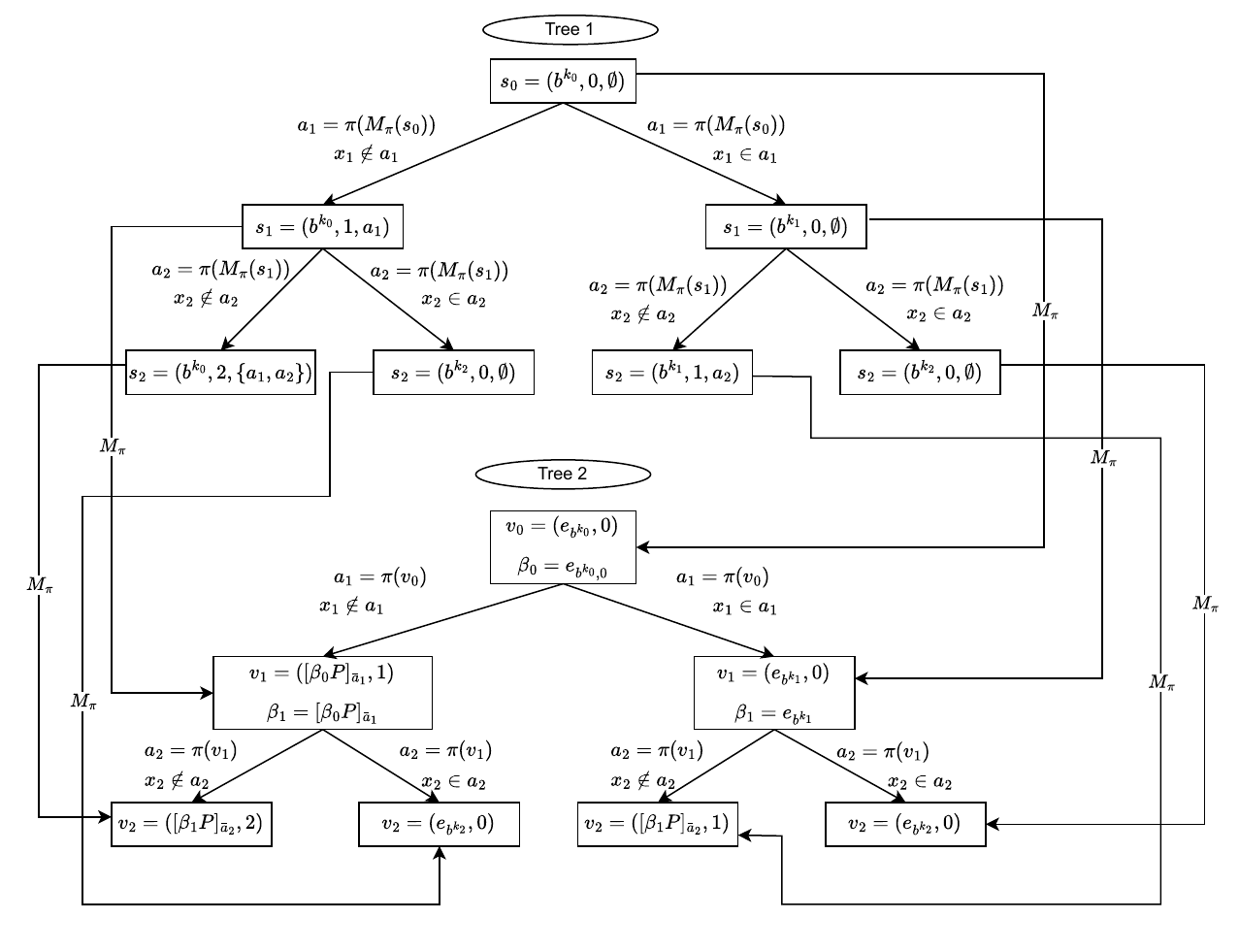}
\caption{For time steps $k=0,1,2$, the upper tree illustrates the progression of beliefs within the POMDP framework, while the lower tree delineates the evolution of states in the Track-MDP. The mapping $M_{\pi}$ between Track-MDP states and POMDP beliefs, is denoted by arrows labeled as $M_{\pi}$. }
\label{fig:mpi_mapping}
\end{figure*}
\subsection{Mapping Beliefs Traversed by POMDP to Track-MDP States}~\label{sec:mapping_equivalence}
\begin{lemma}\label{lem:mapping_lemma}
    Let the state of the Track-MDP at time $k$ be $ s_k = (b^{\tilde{k}},n,h_k)$, and the belief of the POMDP at time $k$ be denoted as $v_k$. Let $v_k$ and $s_k$ be such that $v_k = M_{\pi}(s_k)$. If the Track-MDP transitions to the state $s_{k+1}$ after executing the sensing action $\tilde{\pi}(s_k)$ and the POMDP transitions to the belief $v_{k+1}$ after taking the sensing action $\pi(v_k)$, then it follows that $v_{k+1} = M_{\pi}(s_{k+1})$. Furthermore, $\Tilde{\pi}(s_k) = \pi(v_k)$.
\end{lemma}
\begin{proof}
    First we show that $\tilde{\pi}(s_k) = \pi(v_k)$. From~\eqref{eq:induced_policy},
\begin{align}
       \tilde{\pi}(s_k) = \pi(M_{\pi}(s_k)). 
\end{align}
Since $v_k = M_{\pi}(s_k)$, the equality $\Tilde{\pi}(s_k) = \pi(v_k) $ holds. Let $a_{k+1} = \Tilde{\pi}(s_k) = \pi(v_k) $. Upon executing the sensing action, there are two possibilities, either $x_{k+1} \in a_{k+1}$ or $x_{k+1}\notin a_{k+1}$. 

\noindent\textbf{Case 1:} [$x_{k+1} \in a_{k+1}$]. If $x_{k+1} \in a_{k+1}$, the Track-MDP transitions to the state $s_{k+1} = (x_{k+1},0,\emptyset)$ and the POMDP transitions to the belief $v_{k+1} = (e_{x_{k+1}},0)$. From the definition of $M_{\pi}$, 
\begin{align}~\label{eq:lem_map_eq1}
    (e_{x_{k+1}},0)=M_{\pi}((x_{k+1},0,\emptyset)).
\end{align}
\noindent\textbf{Case 2:} [$x_{k+1} \notin a_{k+1}$]. If $x_{k+1} \notin a_{k+1}$, the Track-MDP transitions to the state $s_{k+1} = (b^{\tilde{k}},n+1,\{h_k,a_{k+1}\} )$ where $a_{k+1} = \pi(v_k)$ and the POMDP transitions to the belief $v_{k+1} = ([\beta_k P]_{\overline{\pi(v_k)}},n+1)$. From the definition of the mapping $M_{\pi}$ in \eqref{eq:mpi_3}, 
\begin{align}~\label{eq:lem_map_eq2}
   ([\beta_k P]_{\overline{\pi(v_k)}} ,n+1)=M_{\pi}((b^{\tilde{k}},n+1,\{h_k,a_{k+1}\} )).
\end{align}
From \eqref{eq:lem_map_eq1} and \eqref{eq:lem_map_eq2}, $v_{k+1}= M_{\pi}(s_{k+1})$.
\end{proof}
As an implication of the lemma, the evolution of the states of the Track-MDP and beliefs of the POMDP can be mapped by $M_{\pi}$ as is evident from the following arguments.
\begin{itemize}
    \item At time $k=0$, the position $x_0 = b^{k_0}$ is known without ambiguity to both the POMDP and the Track-MDP due to the high power sensing action~\ref{sec:assumption_constraint}. As a consequence, the Track-MDP state is $s_0=(b^{k_0},0,\emptyset)$ and the POMDP belief is $v_0 = (e_{b^{k_0}},0)$. The function $M_{\pi}$ maps the Track-MDP state $(b^{k_0},0,\emptyset)$ to the belief $(e_{b^{k_0}},0)$. 
    \item At time $k=1$, the target transitions from $b^{k_0}$ according to the transition matrix $P$ defined in \eqref{eq:process_st_transition}. Let the position of the target $x_1$ be $b^{k_1}$. To detect the target, the POMDP policy turns on sensors in the grid according to action $ \pi(v_0)$ and the Track-MDP turns on sensors in the grid according to $\tilde{\pi}(s_0)$. Since $v_0 =M_{\pi}(s_0)$, by Lemma~\ref{lem:mapping_lemma}, $\tilde{\pi}(s_0) =  \pi(v_0)$. Let $a_1 = \tilde{\pi}(s_0) $.
    \item If the target is detected by the sensing action $a_1$~(i.e., $x_1 \in a_1$), the belief of the POMDP evolves to $v_1 = (e_{b^{k_1}},0)$ and the Track-MDP state evolves to $ s_1 = (b^{k_1}, 0,\emptyset)$. 
    \item Alternatively, if the target is not sensed by the sensing action $a_1$ (i.e., $x_1 \notin a_1$), the belief of the POMDP at $k=1$ evolves to $v_1 = ([e_{b^{k_0}} P]_{\bar{a}_1},1)$ and the state of the Track-MDP evolves to $s_1 = (b^{k_0},1,a_1)$.
    \item In both the cases where the target is detected and undetected, $v_1 = M_{\pi}(s_1)  $ by Lemma~\ref{lem:mapping_lemma}.
    \item In general, at time $k=n$, let the Track-MDP state $s_n$  be mapped by $M_{\pi}$ to the belief $v_n$. From Lemma~\ref{lem:mapping_lemma}, $\pi(v_n) = \tilde{\pi}(s_n)$. Upon executing the sensing action $ \pi(v_n)$ and $ \tilde{\pi}(s_n)$ on the POMDP and the Track-MDP respectively, the belief and the state evolves such that  $v_{n+1} = M_{\pi}(s_{n+1}) $. 
    \item Alternatively, at time $k=n$, if the target has been unobserved for $T_{\mathrm{max}}+1$ times steps by the Track-MDP policy $\tilde{\pi}$, it is unobserved by the POMDP policy $\pi$ as well for $T_{\mathrm{max}}+1$ time steps since $\pi(v_m) = \tilde{\pi}(s_m)$ for all $m\leq n$ by Lemma~\ref{lem:mapping_lemma}. Hence, at time $k=n+1$, both the POMDP and Track-MDP executes the high power sensing action to detect the target. If $x_{n+1} = b^{k_{n+1}}$, the Track-MDP state evolves to $s_{n+1} = (b^{k_{n+1}},0,\emptyset)$ and the POMDP state evolves to $v_{n+1} = e_{b^{k_{n+1}}}$. By the definition of the mapping, $v_{n+1}=M_{\pi}(s_{n+1})$.
\end{itemize}
Hence, from the definition of $M_{\pi}$ and the induced policy $\tilde{\pi}$, it is evident that the states traversed by the Track-MDP under the policy $\tilde{\pi}$ can be mapped to the beliefs traversed by the POMDP policy $\pi$ while tracking the target, and the actions taken by both the policies at each time step $k$ are the same. Since, the sensing actions are the same, the observations are identical at each time step $k$. Hence the reward $r_k$, which is a function of action and observation~(as defined in~\eqref{eq:reward_func},\eqref{eq:inf_hori_rew},\eqref{eq:pomdp_value}) at each time step $k$ is identical for both the POMDP policy $\pi$ and the Track-MDP policy $\Tilde{\pi}$.  
\begin{remark}
  The evolution of the states and beliefs for time $k=0,1,2$ and the mapping $M_{\pi}$ are described in Figure~\ref{fig:mpi_mapping}.  
\end{remark}
\subsection{Implication of Mapping and Induced Policy}
Consider the optimal POMDP policy $\pi^{*}_{\mathrm{POMDP}}$. Let $\tilde{\pi}^{*}_{\mathrm{POMDP}}$ be the policy induced on the Track-MDP by the map $M_{\pi^{*}_{\mathrm{POMDP}}}$ as described in Section~\ref{sec:induced_policy}. From the discussion in Section~\ref{sec:mapping_equivalence} and Lemma~\ref{lem:mapping_lemma}, the states traversed on the Track-MDP by the policy $\tilde{\pi}^{*}_{\mathrm{POMDP}}$ can be mapped to beliefs traversed by the POMDP with the policy $\pi^{*}_{\mathrm{POMDP}}$ and the rewards $r_k$ are the same for all $k\geq 0$. Hence, it is easy to show that the $\mathrm{IHTR}$ of the Track-MDP policy $\tilde{\pi}^{*}_{\mathrm{POMDP}}$ is the same as the $\mathrm{IHTR}$ of the optimal POMDP policy. We state this result formally as Theorem~\ref{thm:optimalaspomdp} {in Section~\ref{sec:theory}}.
\begin{remark}
    The policy $\tilde{\pi}^{*}_{\mathrm{POMDP}}$ is optimal for the Track-MDP since the IHTR of any other policy on the Track-MDP cannot be greater than the IHTR of the optimal POMDP policy $\pi^{*}_{\mathrm{POMDP}}$. This is the case since the Track-MDP is a surrogate MDP for solving for the best policy in terms of IHTR on the POMDP for the {TTCS} problem.
\end{remark}
\begin{remark}
    The mapping $M_{\pi}$ and Lemma~\ref{lem:mapping_lemma} apply for all $T_{\mathrm{max}}>0$, ensuring consistency between Track-MDP states and POMDP beliefs, even as $T_{\mathrm{max}}$ approaches infinity. The state space of the Track-MDP is countably infinite and the belief space of the POMDP is uncountable and infinite. Despite this vast difference in cardinality, the $\mathrm{IHTR}$ of $\pi^{*}_{\mathrm{TMDP}}$ and $\pi^{*}_{\mathrm{POMDP}}$ are the same. Additionally, an exact mapping $ M_{\pi}$ is only possible for the sensing model defined in Section~\ref{sec:sensing_grid}. Other models may only allow approximations.
\end{remark}
\begin{remark}
For the energy-efficient {TTCS problem}, learning the optimal policy in the POMDP formulation, denoted as $\pi^*_{\mathrm{POMDP}}$, is highly complex due to the uncountably infinite set of possible beliefs $v_k$~\cite{ross2008online}. Further, even if the motion model of the target~($P$) was completely known, finding the optimal POMDP policy is PSPACE-complete~\cite{papa_tsitsiklis}. Hence, sub-optimal solutions like $\mathrm{Q}_{\mathrm{MDP}}$ and First Cost Reduction ($\mathrm{FCR}$) are proposed alternatives.
Our study suggests using RL with the Track-MDP framework to find the optimal sensing policy, offering a simpler alternative to directly solving for $\pi^*_{\mathrm{POMDP}}$. Learning the optimal policy within an MDP framework is notably easier~\cite{littman_complexity}\cite{ucrl}, and aligns with the optimal POMDP policy, indirectly enabling its learning.
\end{remark}
\section{Theoretical Results}\label{sec:theory}
In this section, we study the theoretical properties of the optimal RL policy $\pi^*_{\mathrm{TMDP}}$ on the Track-MDP.  
\subsection{Tracking Guarantee}
The optimal policy $\pi^*_{\mathrm{TMDP}}$ in Track-MDP maximizes the sum of pseudo-rewards $R(x_k,a_{k+1})$ over time, which depends on parameters $c$ and $r$. The values of $c$ and $r$ significantly impact the pseudo-reward's behavior. For instance, when $c=0$ and $r\neq 0$, the rewards encourage active tracking without penalizing sensor activation, leading to a policy that activates all sensors to locate the target with certainty. Conversely, with $c\neq 0$ and $r=0$, the policy tends to deactivate sensors and miss the target, resulting in low target tracking probability. This correlation between $c$ and $r$ values and tracking probability is formally defined as the Track Property.
\begin{theorem}[Track Property]
\label{tracking_property}
Let the state of the Track-MDP at time $k$ be  $s_k=(x_{l(k)},n(k),h_k)$. For the process Markov chain $(X_k)_{k\geq 0}$ with state transition probability ${P}$, all grid positions ${b^j}\in \mathcal{X}$ such that $\mathrm{Pr}(X_{k+1} = {b^j} | s_k)> \frac{c}{r}$ will have sensors turned on by the optimal policy $\pi^*_{\mathrm{TMDP}}$.
\end{theorem}
\subsubsection*{Implication of Track Property}
The theorem guarantees that when the predicted probability
of the target being located in a cell of the grid is greater than
$\frac{c}{r}$, then the optimal policy will turn on a sensor in that cell and track the target. This justifies the reward model considered in this work. The proof of the theorem is presented in the Appendix~\ref{proof:tracking_property}. 
\subsection{State Estimator}\label{subsec:estimator}
The optimal estimate of the position when the observation $y_k = x_k$ is the observation $y_k$. However, in the case where $y_k = \mathcal{E}$, the state of the target $x_k$ becomes uncertain. Hence, we derive an estimate of the position of the target $\hat{X}_k$ from the $Q$ function~\cite{sutton2018reinforcement} of policy $\pi^*_{\mathrm{TMDP}}$.
\begin{theorem}\label{thm:estimator}
    Let $\pi^*_{\mathrm{TMDP}}$ be the optimal policy on the Track-MDP and $Q^{\pi^*_{\mathrm{TMDP}}}$ be the Q-value of the optimal policy on the Track-MDP. 
    Let the state of the Track-MDP at time $k$ be $s_k = (x_{l(k)},n(k),h_k)$. Let $a_k^* = \pi^*_{\mathrm{TMDP}}(s_k)$ and $y_{k+1} = \mathcal{E}$. Consider the estimator 
\begin{equation*}
    \hat{X}_{k+1} = \arg \max_{x \in \mathcal{X} , x \notin a_k^*} \frac{Q^{\pi^*_{\mathrm{TMDP}}}(s_k, \tilde{a})  - Q^{\pi^*_{\mathrm{TMDP}}}(s_k, \tilde{a}_x) + c}{r}.
\end{equation*}
The estimator $\hat{X}_{k+1}$ is the Maximum A Posteriori~(MAP) estimator of the state $X_{k+1}$ given the current state $s_k$ of the Track-MDP. The proof of the theorem is presented in the Appendix~\ref{proof:estimator}. 
\end{theorem}

%
\subsection{Comparison of IHTR of \texorpdfstring{$\pi^*_{\mathrm{POMDP}}$}{TEXT} and \texorpdfstring{$\pi^*_{\mathrm{TMDP}}$}{TEXT} }\label{subsec:perf_bounds}
The beliefs traversed by POMDP policy $\pi^*_{\mathrm{POMDP}}$ while tracking the target can be mapped to the states traversed by the induced Track-MDP policy $\tilde{\pi}^*_{\mathrm{POMDP}}$. As a consequence, we prove the following theorem. The proof is presented in the Appendix~\ref{proof:optimalaspomdp}. 
\begin{theorem}\label{thm:optimalaspomdp}
     Let $\pi^{*}_{\mathrm{TMDP}}$ be an optimal Track-MDP policy and $\pi^{*}_{\mathrm{POMDP}}$ be the optimal POMDP policy for tracking a target with transition matrix $P$. Then,     
\begin{equation*}
    V^{\pi^{*}_{\mathrm{TMDP}}}_{\mathrm{TMDP}}((b^j,0,\emptyset)) = V^{\pi^{*}_{\mathrm{POMDP}}}_{\mathrm{POMDP}}((e_{b^j},0))
\end{equation*}
for all $b^j \in \{b^i\}_{i=1}^{N^2}$.
\end{theorem}
\section{Experiments} \label{sec:exp}
An $N\times N$ sensing grid is considered, as outlined in Section~\ref{sec:sensing_grid}, with $N=10$, and each cell is equipped with a sensor. We employ the POMDP policy $\pi_{\mathrm{Q}_{\mathrm{MDP}}}$~\cite{littman_qmdp} due to the computational intractability  of $\pi^*_{\mathrm{POMDP}}$. If the belief at time $k$ is $v_k = (v_k,n(k))$, the $\mathrm{Q}_{\mathrm{MDP}}$ policy $\pi_{\mathrm{Q}_{\mathrm{MDP}}}$ is 
\begin{align}\label{eq:qmdp_policy}
\pi_{\mathrm{Q}_{\mathrm{MDP}}}(v_k)=
\begin{cases}
\{b^j: [\beta_k P]_j \geq \frac{c}{r},\ j\in [N]^2]\}, &\hspace{-0.75em} n(k)\leq T_{\max} \\
\hat{a}_s, & \hspace{-3em} n(k) = T_{\max}+1
\end{cases}.
\end{align}
The derivation of \eqref{eq:qmdp_policy} is presented in the Appendix~\ref{appendix:qmdp}. The reward parameters are $r=1$ and $D=N^2 c$. We illustrate the average infinite horizon tracking reward~(defined in Section~\ref{sec:aihtr}) for $c$ values  $0.16, 0.18, 0.20, 0.22$.

The target has a transition kernel $P$ such that, if the target is at position $x_k$, at time $k$, then at time ${k+1}$, the target moves to one of either $3$, $4$ or $5$~(defined as parameter $Z=3,4,5$) random positions in a $3\times 3$ grid around the current position. For example, if $x_k = b^{13}$ on the sensing grid, shown in Figure~\ref{code1}, then in the $3\times3$ region around $b^{13}$, the target moves randomly to $4$ of the $9$ positions if $Z=4$. Let the positions be $b^9,b^{12},b^{18},b^{19}$. The transition probability $P$ is defined to allow the target to exit the grid with probability $P(x_{k+1} = b^T|x_{k}) =0.005$, move to one of the $Z$ cells, say $b^9$, with probability $P(x_{k+1}= b^9|x_k = b^{13}) =0.15$, and transition to the remaining cells $b^{12},b^{18},b^{19}$ with probability $P(x_{k+1}|x_k = b^{13}) = (1-0.15-0.005)/(Z-1)$~(i.e., uniformly distributed among the remaining cells).

We set the transition probability $p(x_{k+1}|x_k)$ as $0.15$ for one of the states $x_{k+1}$, since $0.15 < \frac{c}{r}$ for all $c$ values considered. To further elaborate our choice of parameter, consider the case when $c=0.14$ and $Z=4$. Then, $0.15 > \frac{c}{r}$ and also $ (1-0.15-0.005)/(Z-1) > \frac{c}{r}$. Hence, by~\eqref{eq:qmdp_policy} and Theorem~\ref{tracking_property}, both the $\mathrm{Q}_{\mathrm{MDP}}$ and the Track-MDP will turn on all the sensors exactly in the support of $P(x_{k+1}|x_{k})$, thus both tracking policies will have the same reward and $100\%$ tracking accuracy. This is not an interesting case to study and hence we simulate the alternative case where at least one of the probabilities in the support of $p(x_{k+1}|x_k)$ does not satisfy the condition $\mathrm{Pr}(X_{k+1} = {b^j} | s_k)> \frac{c}{r}$ of Theorem~\ref{tracking_property}. 
\subsection{Average Infinite Horizon Tracking Reward}\label{sec:aihtr}
\begin{figure*}%
\centering
\begin{subfigure}{0.41\columnwidth}
\includegraphics[width=\columnwidth]{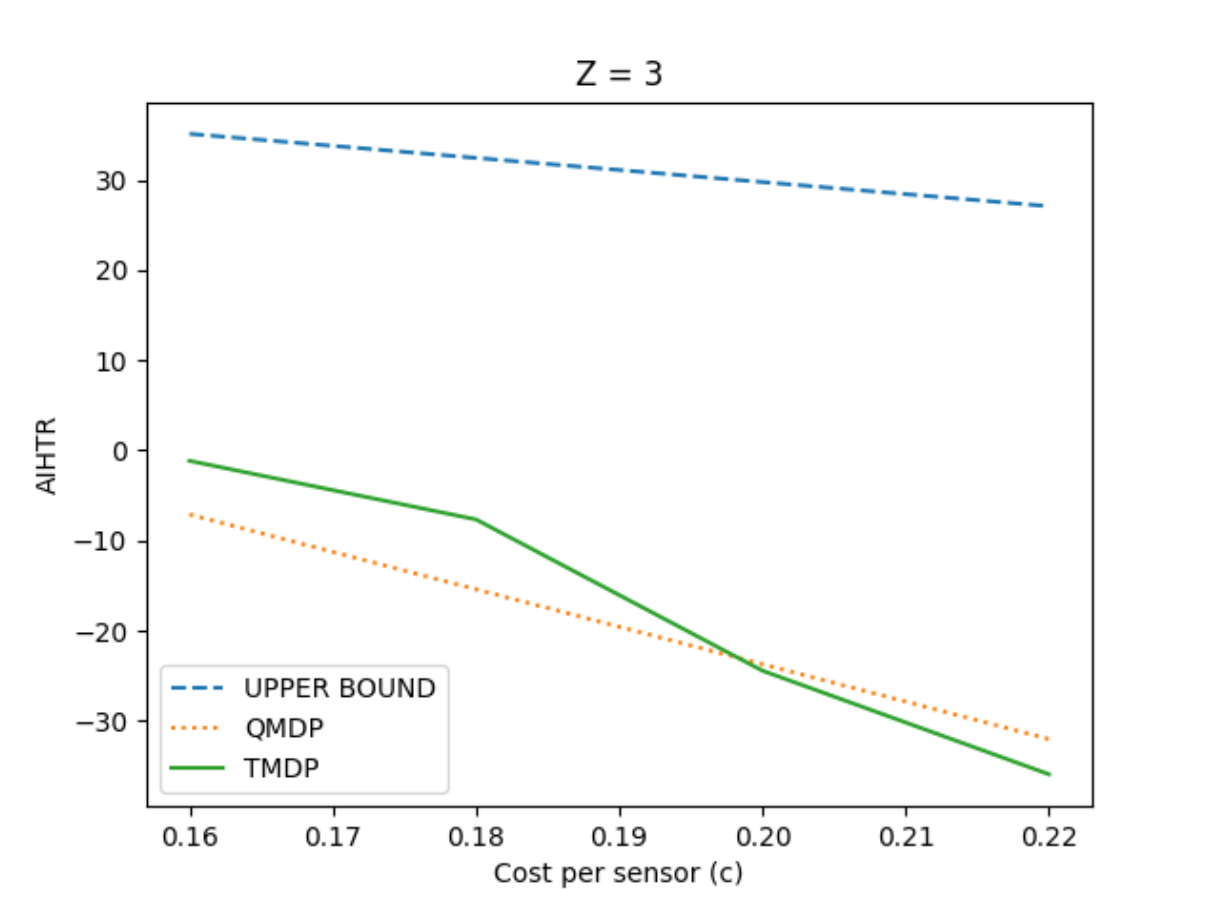}%
\caption{\scriptsize AIHTR vs cost per sensor when $Z=3$ }%
\label{z_3}%
\end{subfigure}\hfill%
\begin{subfigure}{.41\columnwidth}
\includegraphics[width=\columnwidth ]{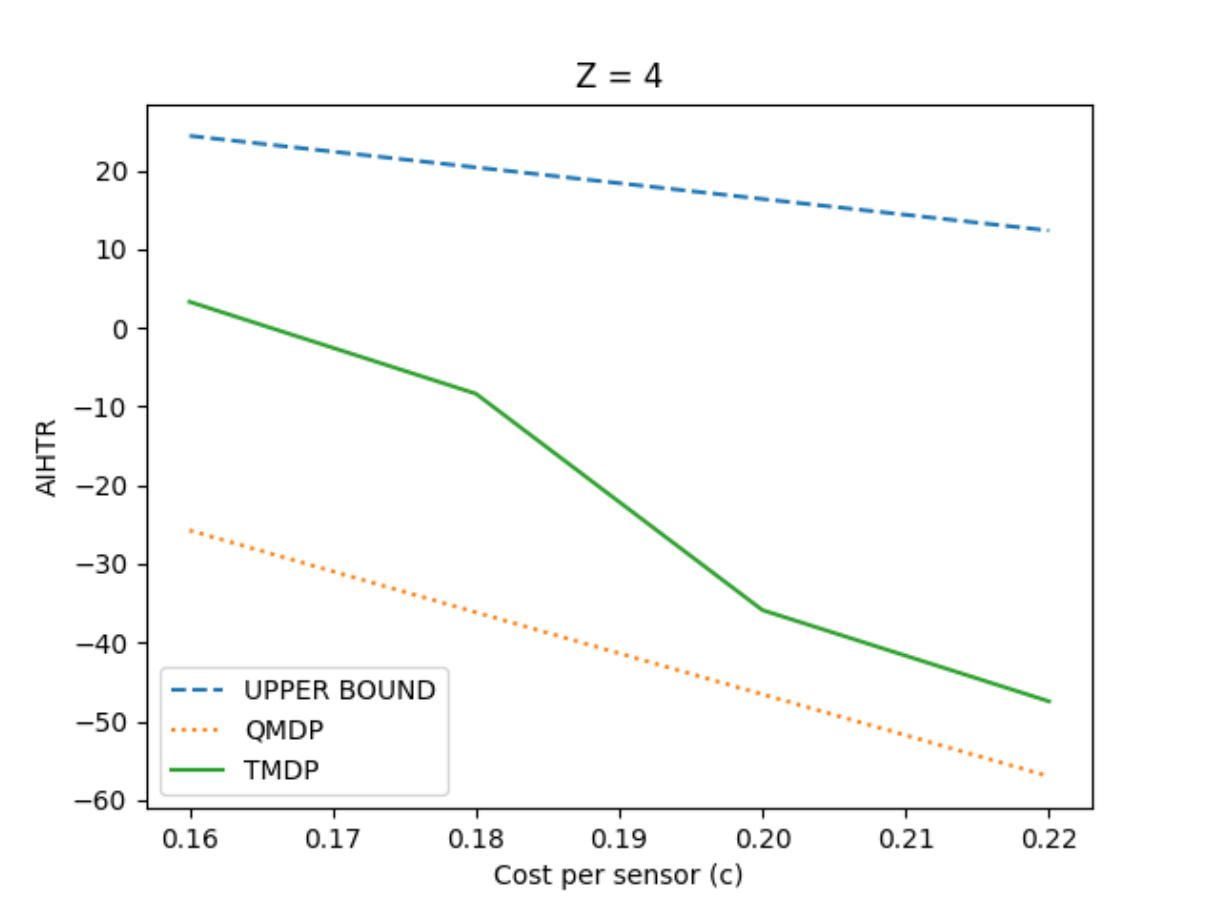}%
\caption{{\scriptsize AIHTR vs cost per sensor when $Z=4$}}%
\label{z_4}%
\end{subfigure}\hfill%
\begin{subfigure}{.41\columnwidth}
\includegraphics[width=\columnwidth ]{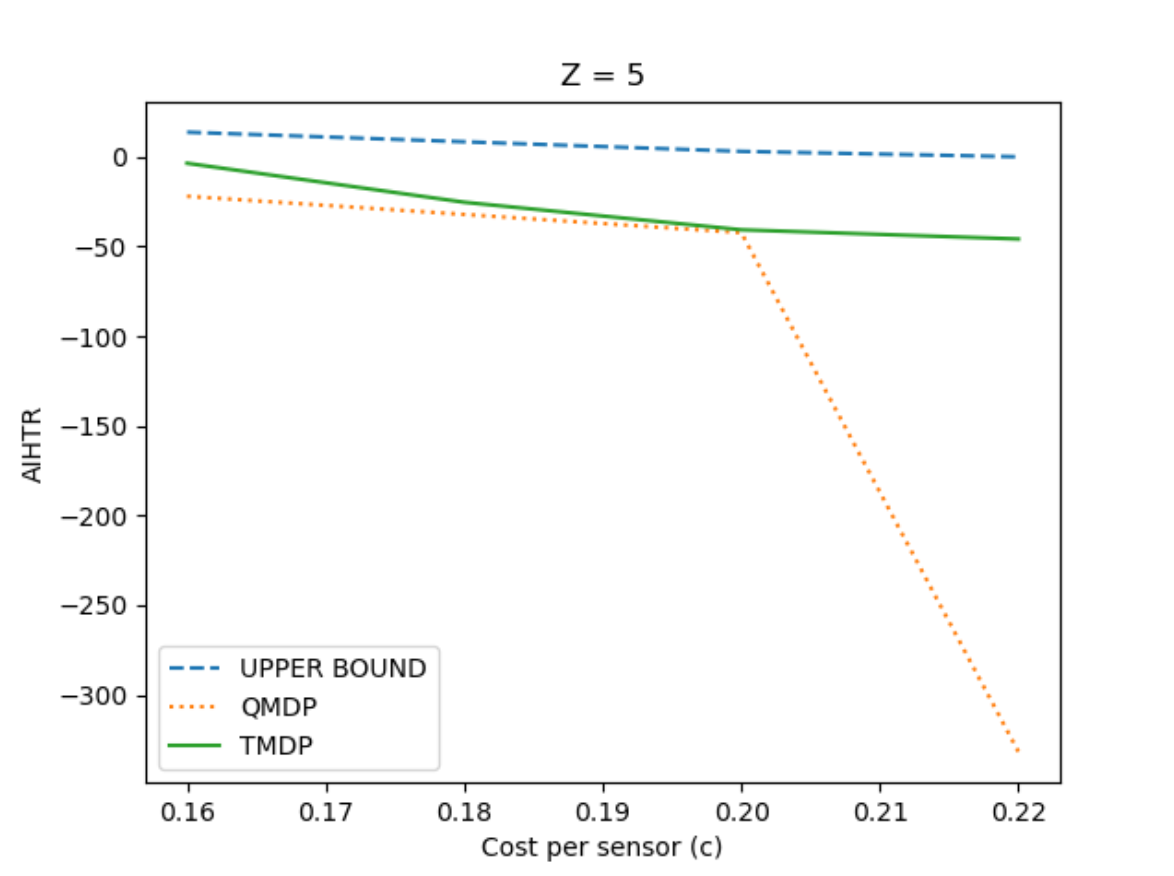}%
\caption{{\scriptsize AIHTR vs cost per sensor when $Z=5$}}%
\label{z_5}%
\end{subfigure}\hfill%
\caption{AIHTR vs cost per sensor~($c$) for cardinality of support $Z$ of $P(x_{k+1}|x_k)$ values $3,4,5$.}
\end{figure*}

 We compare the Average Infinite Horizon Tracking Reward~(AIHTR) of the Track-MDP~(denoted as TMDP), $\mathrm{Q}_{\mathrm{MDP}}$ policy and the upper bound $V^*_{\mathrm{Q}_{\mathrm{MDP}}}$, described in Section~\ref{sec:upper_bound} of the Appendix~\ref{sec:upper_bound}. The AIHTR is defined as 
\begin{align}
    \mathrm{AIHTR}_{\mathrm{TMDP}} &= \frac{\sum_{j=1}^{N^2} V^{\pi^*_{\mathrm{TMDP}}}_{\mathrm{TMDP}}((b^j,0,\emptyset))}{N^2} &&\\
    \mathrm{AIHTR}_{\mathrm{Q}_{\mathrm{MDP}}} &= \frac{\sum_{j=1}^{N^2} V^{\pi_{\mathrm{Q}_{\mathrm{MDP}}}}_{\mathrm{POMDP}}(e_{b^j})}{N^2} &&\\
    \mathrm{AIHTR}_{\mathrm{Upper\:bound}} &= \frac{\sum_{j=1}^{N^2} V^{*}_{\mathrm{Q}_{\mathrm{MDP}}}(e_{b^j})}{N^2}.
\end{align}
In Figures~\ref{z_3}~\ref{z_4}~\ref{z_5}, we plot the AIHTRs for $Z$ values $3$, $4$ and $5$, respectively, averaged over $1000$ episodes. The policy $\pi^*_{\mathrm{TMDP}}$ is estimated with Advantage Actor Critic~(A2C)~\cite{a2c} and $V^{\pi^*_{\mathrm{TMDP}}}_{\mathrm{TMDP}}$ and $V^{\pi_{\mathrm{Q}_{\mathrm{MDP}}}}_{\mathrm{POMDP}}$ are evaluated empirically. The AIHTR of the Track-MDP is consistently higher than the $\mathrm{Q}_{\mathrm{MDP}}$. This is despite the fact that $\mathrm{Q}_{\mathrm{MDP}}$ has complete knowledge of the target's dynamics $P$, while the Track-MDP is model agnostic. 
\begin{remark}
    We plot the upper bound $V^*_{\mathrm{Q_{MDP}}}$, since the value of $V_{\mathrm{POMDP}}^{\pi^*_{\mathrm{POMDP}}}$ is difficult to compute.
\end{remark}
\begin{remark}
    In Figure~\ref{z_5}, a large drop can be observed in $\mathrm{AIHTR}$ for the $\mathrm{Q_{MDP}}$ policy when $c =0.22$. This is due to the fact that $[\beta_k P]_j < \frac{c}{r}$ for all positions $b^j$ in the grid and all reachable beliefs $\beta_k$. As a consequence, $\pi_{\mathrm{Q_{MDP}}}$ does not turn on any sensors in the grid except when ${n(k)= T_{\mathrm{max}} + 1}$.
\end{remark}
\subsection{Accuracy Comparison}
In this section, we compare the Hamming tracking accuracy of Track-MDP with $\mathrm{Q}_{\mathrm{MDP}}$. Hamming accuracy measures the fraction of perfect observations~($y_k = x_k$) in an episode~(time between target entering and exiting the grid). While Track-MDP's optimal policy implicitly learns target state transitions~(Theorem~\ref{thm:estimator}), evaluating it based on Hamming accuracy minimally uses this knowledge,
since an estimation accuracy of 0 is assumed when the target
is missed. According to Table~\ref{table_2}, Track-MDP outperforms $\mathrm{Q}_{\mathrm{MDP}}$ across various sensor cost values, $c$, demonstrating its advantage in achieving high tracking accuracy without relying on a model.

\begin{table}[h]
\centering
\scalebox{1}{
\begin{tabular}{c|c|c|c}
\textbf{Z} & $\mathbf{c}$     & $\mathrm{Q}_{\mathrm{MDP}}$ \textbf{Acc}      & ${\mathrm{TMDP}}$ \textbf{Acc}   \\ \hline
\textbf{5}   & 0.16 & 89.95\% & \textbf{93.75\%}   \\ \hline
\textbf{4}   & 0.16 & 88.82\% & \textbf{94.28\%}   \\ \hline
\textbf{3}   & 0.16 & 90.74\% & \textbf{92.39\%} \\ \hline
\textbf{5}   & 0.18 & 89.95\% & \textbf{94.94\%} \\ \hline
\textbf{4}   & 0.18 & 88.82\% & \textbf{95.56\%} \\ \hline
\textbf{3}   & 0.18 & 90.74\% & \textbf{92.06\%} \\ \hline
\textbf{5}   & 0.20 & 89.95\% & \textbf{92.45\%} \\ \hline
\textbf{4}   & 0.20 & 88.82\% & \textbf{89.7\%} \\ \hline
\textbf{3}   & 0.20 & 90.74\% & \textbf{93.23\%} \\ \hline
\textbf{5}   & 0.22 & 54.3\% & \textbf{90.21\%} \\ \hline
\textbf{4}   & 0.22 & 88.82\% & \textbf{88.85\%} \\ \hline
\textbf{3}   & 0.22 & \textbf{90.74}\% & {87.18\%} \\ \hline
\end{tabular}
}
\caption{Hamming accuracy of Track-MDP vs $\mathrm{Q}_{\mathrm{MDP}}$ for sensor cost $c$ and cardinality of support $Z$ of $P(x_{k+1}|x_k)$.}
\label{table_2}
\end{table}
\section{Conclusion}
In our work, we introduce the Track-MDP, an RL-based approach to {the TTCS problem} that circumvents the partial observability issue without relying on POMDP methods. We demonstrate the tracking capabilities of the optimal Track-MDP policy and establish its equivalence to the optimal POMDP policy in terms of infinite-horizon tracking rewards. Through simulations, we show that our Track-MDP policy consistently outperforms the $\mathrm{Q}_{\mathrm{MDP}}$ policy in both AIHTR and Hamming tracking accuracy across a range of sensor costs. Although we considered a small tracking grid with $100$ sensors in our experiments, the proposed method can be extended to larger grids, by partitioning into smaller sub-grids. Transitioning between sub-grids can be facilitated through a hand-off algorithm, which estimates and activates sensors in adjacent grids when necessary. The Track-MDP framework can be extended to more realistic target motion and sensing models~\cite{BarShalom2001EstimationWA,BarShalom_handbook}, which we defer to future work. 
\section{Acknowledgement}
We would like to thank Prof. Bhaskar Krishnamachari for helpful discussions regarding the equivalence of the POMDP and Track-MDP models.
\bibliographystyle{IEEEtran}
\bibliography{trackmdp}
\section{Appendix}
\begin{definition}[Valid paths]
    For the stationary Markov Chain $X_k$ of the process model, define the set of valid paths of length $L$ as
\begin{align*}
\mathcal{Z}^{L}=\{ (g_0&,\cdots,g_{L-1}): g_i \in \mathcal{X}, &&\\
\nonumber&P(X_{1}=g_1,\cdots,X_{L-1}=g_{L-1}|X_0=g_0)>0 \}   
\end{align*}
\end{definition}
\begin{definition}[Valid unobserved paths under action tuple] Consider $\mathcal{Z}^{L}$, the set of valid paths of length $L$ of the stationary Markov Chain $X_k$. Let $\zeta^{L}\in \mathcal{Z}^L$, where $\zeta^{L}(j)= g_j$. Consider a $(L-2)$-element tuple of actions 
\begin{equation*}
v^{L-2}=(v_0,\cdots,v_{L-3}),~\text{i.e. }\quad v^{L-2}[i]=v_i, 0\leq i\leq L-3   
\end{equation*}
where $v_i \in \mathcal{A}$, the action space of Track-MDP. The set of valid unobserved paths of length $L$ under the action tuple $v$ is defined as 
\begin{equation*}
\bar{\mathcal{Z}}^L_{v}=\{\zeta^L \in \mathcal{Z}^L: \zeta^L(i+1)\notin v^L[i], 0\leq i \leq L-3\},
\end{equation*}  
\end{definition}
\begin{definition}[Valid unobserved paths under action tuple with origin]  We define the set of valid unobserved paths, i.e., $\bar{\mathcal{Z}}^{L}_{v}$, of length $L$ and originating from $b\in \mathcal{X}$ by
\begin{equation*}   
    \mathcal{T}(b,L,v)=\{\bar{\zeta}^{L} : \bar{\zeta}^{L} \in \bar{\mathcal{Z}}^{L}_{v}, \bar{\zeta}^{L}(0)=b \}.
 \end{equation*}   
\end{definition}
\begin{definition}
The probability that the state $X_{k+1}$ equals $\bar{b}\in \mathcal{X}$ conditioned on $s_k = (x_{l(k)},n(k),h_k)$ is defined as
\begin{equation*}
\Bar{P}(\bar{b},s_k) \triangleq \mathrm{Pr}(X_{k+1} =\bar{b}| s_k) =   \frac{P(\bar{b},s_k)}{\sum_{\bar{b}\in \mathcal{B}(s_k)}P(\bar{b},s_k)} 
\end{equation*}
where 
\begin{align*}
    P(\bar{b},s_k)=P\left(\{\bar{\zeta}^{n} : \bar{\zeta}^{n}\in \mathcal{T}(x_{l(k)},n(k)+2,h_{k}):\right. \left.\bar{\zeta}^{n}(n(k)+1)=\bar{b} \}\right),
\end{align*}

\begin{equation*}
    \begin{split}
        \mathcal{G}(\bar{b},s_k)=\{\bar{\zeta}^{n}\in \mathcal{T}(x_{l(k)},n(k)+2,h_{k}): \bar{\zeta}^{n}(n(k)+1)=\bar{b}\},
    \end{split}
\end{equation*}

\begin{equation*}
 \begin{split}   
\mathcal{B}(s_k)=\{ \bar{b} \in \mathcal{X}: \mathcal{G}(\bar{b},s_k)\neq \emptyset\}.
 \end{split}   
\end{equation*}
\end{definition}
\begin{definition}
    Define the operator $\hat{T}:\mathcal{S} \rightarrow \mathcal{S}$ mapping states of Track-MDP. For a state $s =(x_{l(k)},n(k),h_k)$, with $n(k)\leq T_{\max}$ and an action $a \in \mathcal{A}$,  
    $$\hat{T}(s,a) = \hat{T}((x_{l(k)},n(k),h_{k}),a) =  (x_{l(k)},n(k)+1,(h_{k},a)), $$
$$\hat{T}(s,\{\}) = s.$$    
The operator defines the state that the Track-MDP transitions to if its initial state is $s$ and the target is not sensed after the sensing action $a$. Also if $n(k)+n \leq T_{\max} +1$, then with slight abuse of notation, we define for a sequence of actions $a_1,a_2,\cdots,a_n$, 
\begin{align*}
    \hat{T}(s,(a_1,\cdots,a_n) ) = (x_{l(k)},n(k)+n,(h_{k},a_1,\cdots,a_n)).
\end{align*}
\end{definition}
\begin{definition}
 Consider the operator $\hat{T}^n$ defined on the state $s_0= (x_{l(k)},n(k),h_{k}) $, with $n(k)+ n \leq T_{\max}+1$ and $\pi$ be a policy on the Track-MDP. Consider the following sequence of states
 \begin{equation*}
 \begin{aligned}
     s_0 = \hat{T}(s_0, \{\}),
     s_1 = \hat{T}(s_0, \pi(s_0)) ,
        \cdots,
     s_n = \hat{T}(s_{n-1}, \pi(s_{n-1})) &&\\
 \end{aligned}    
 \end{equation*}
 Thus, for compactness we define,
 \begin{equation*}
     s_n = {\hat{T}^n}(s_0,\pi)
 \end{equation*}
 which essentially applies the previous procedure $n$ times starting from $s_0$ and following policy $\pi$.
\end{definition}
To prove the tracking property of Track-MDPs, we first prove the following lemma.
\begin{lemma}\label{lem:tower_bound}
For the Track-MDP with optimal policy $\pi^*_{\mathrm{TMDP}}$ and a process transition model $P$ with Markov process $(X_k)_{k\geq 0}$ and state $s_k= (x_{l(k)},n(k),(h_k))$, at time $k$,
\begin{align}\label{eq:vis_p1}
 \mathbb{E}[V^{\pi^*_{\mathrm{TMDP}}}((x_{l(k)},&n(k)+1,\{h_k,a^*_k\}))|x_{k+1}=\bar{b}] &&\\
 \nonumber &\leq \mathbb{E}[V^{\pi^*_{\mathrm{TMDP}}}((\bar{b},0,\emptyset))|x_{k+1}=\bar{b}]
 \end{align}
 \begin{align}\label{eq:vis_pr2}
 &\mathbb{E}[V^{\pi^*_{\mathrm{TMDP}}}((x_{l(k)},n(k)+1,\{h_k,a^*_k\}))|x_{k+1} \in \mathcal{B}(s_k)\setminus\{a^*_k \cup \bar{b}\}]  &&\\ 
 \nonumber & \leq  \mathbb{E}[V^{\pi^*_{\mathrm{TMDP}}}((x_{l(k)},n(k)+1,\{h_k,a^*_k\cup \bar{b}\}))|x_{k+1} \in \mathcal{B}(s_k)\setminus\{a^*_k \cup \bar{b}\}],
\end{align}

where  $a_k^* = \pi^*_{\mathrm{TMDP}}(s_k)$ and $\Bar{b}\in \mathcal{B}(s_k)$.
\end{lemma}
\begin{proof}
For brevity we show the proof for
\begin{align*}
\mathbb{E}[V^{\pi^*_{\mathrm{TMDP}}}((x_{l(k)},&n(k)+1,\{h_k,a^*_k\}))|x_{k+1}=\bar{b}] &&\\
&\leq \mathbb{E}[V^{\pi^*_{\mathrm{TMDP}}}((\bar{b},0,\emptyset))|x_{k+1}=\bar{b}].
\end{align*}
Define the states $s_{01} = (x_{l(k)},n(k)+1,\{h_k,a^*_k\}) $ and $s_{02} = (\bar{b},0,\emptyset) $ and let the target be at position {$x_{k+1} = b$ } at time step $k+1$. Define state $\hat{s}_n$ and $\hat{a}_n$ as 
\begin{align*}
    {\hat{a}_n = \Big(\pi^*_{\mathrm{TMDP}} ({\hat{T}^0}(s_{01}, \pi^*_{\mathrm{TMDP}})),\cdots, \pi^*_{\mathrm{TMDP}}({\hat{T}^{n-1}}(s_{01}, \pi^*_{\mathrm{TMDP}}))\Big)}
\end{align*}
\begin{align*}
    \hat{s}_n \triangleq  \hat{T}\Big(s_{02},\hat{a}_n \Big)
\end{align*}
Consider two instances of a Track-MDP, with starting states $s_{01}$ and $s_{02}$ respectively. Define a policy $\hat{\pi}$ such that 
\begin{equation*}
\hat{\pi}(s)=
\begin{cases}
\pi^*_{\mathrm{TMDP}}(s_{01}), &if \ s = s_{02} \\
\pi^*_{\mathrm{TMDP}}\Big({\hat{T}^n}(s_{01}, \pi^*_{\mathrm{TMDP}}) \Big), & if \  s = \hat{s}_n, \ n\geq 1\\
\pi^*_{\mathrm{TMDP}}(s) & \ otherwise
\end{cases}
\end{equation*}
Under the new policy $\hat{\pi}$, 
\begin{align*}
 \mathbb{E}[V^{\pi^*_{\mathrm{TMDP}}}((x_{l(k)},&n(k)+1,\{h_k,a^*_k\}))|x_{k+1} 
 =\bar{b}]&&\\
 \nonumber &=  \mathbb{E}[V^{\hat{\pi}}((\bar{b},0,\emptyset))|x_{k+1}=\bar{b}].
\end{align*}
The equality holds due to the sequence of rewards obtained by following policy $ \pi^*$ from $s_{01}$ and and following $\hat{\pi}$ from $s_{02}$ are the same.
Due to the optimality of $\pi^*$, 
\begin{align*}
 \mathbb{E}[V^{\hat{\pi}}&((\bar{b},0,\emptyset))|x_{k+1}=\bar{b}] \leq  \mathbb{E}[V^{\pi^*_{\mathrm{TMDP}}}((\bar{b},0,\emptyset))|x_{k+1}=\bar{b}].
\end{align*}
This proves inequality~\eqref{eq:vis_p1}. By a similar construction and argument, the other inequality in the Lemma  can be proved.
\end{proof}
\subsection{Proof of Theorem~\ref{tracking_property}} \label{proof:tracking_property}
\begin{proof}
Let $s_k=(x_{l(k)},n(k),h_k)$ and the optimal policy be denoted by $\pi^*_{\mathrm{TMDP}}$. \textit{With a slight abuse of notation we use $\pi^*$ instead of $\pi^*_{\mathrm{TMDP}}$ in this proof}. Here $\gamma=1$, but a discount factor implicitly appears due to the target exiting the grid. 
Let $\bar{b}$ be a state in $\mathcal{B}(s_k)$, such that $\bar{P}(\bar{b},s_k)r-c>0$ and $\bar{b} \notin a^*_k$,
\begin{equation}\label{eq:m1}
\begin{aligned}
V^{\pi^*}(s_k)&=\mathbb{E}[R(s,a)+\gamma V^{\pi^*}(S_{k+1}=s')|s=s_k,\: a\sim \pi^*] &&  \\
&=\mathbb{E}[R(s_k,a_k^*)] &&\\
& \hspace{0.2em}+\gamma \sum_{s'}P(S_{k+1}=s'|S_{k}=s_k,a_k^*=\pi^*(s_k))V^{\pi^*}(s') && \\
&={\sum_{b^j\in a^*_k}(\bar{P}(b^j,s_k)r-c)+\gamma \sum_{b^j \in a^*_k}\bar{P}(b^j,s_k)V^{\pi^*}((b^j,0,\emptyset))} &&\\
&\hspace{0.5em}{+\gamma \sum_{b^j \notin a^*_k}\bar{P}(b^j,s_k)V^{\pi^*}((x_{l(k)},n(k)+1,\{h_k,a^*_k\}))}&&\\
&<\bar{P}(\bar{b},s_k)r-c+\sum_{b^j\in a^*_k}(\bar{P}(b^j,s_k)r-c) &&\\
&\hspace{0.5em}+\gamma \sum_{b^j \in a^*_k}\bar{P}(b^j,s_k)V^{\pi^*}((b^j,0,\emptyset))
&&\\
&\hspace{0.5em} +{\gamma \sum_{b^j \notin a^*_k}\bar{P}(b^j,s_k)V^{\pi^*}\left((x_{l(k)},n(k)+1,\{h_k,a^*_k\})\right)}
\end{aligned}
\end{equation}
From the law of total expectation, 
\begin{equation}\label{eq:tower_equation}
\begin{aligned}
&V^{\pi^*}((x_{l(k)},n(k)+1,\{h_k,a^*_k\})) &&\\
& =\frac{\Bar{P}(\bar{b},s_k)}{\sum_{b\in \mathcal{B}(s_k)\cap (a^*_k)^c} \Bar{P}(b,s_k)} &&\\
& \hspace{3em}\mathbb{E}[V^{\pi^*}((x_{l(k)},n(k)+1,\{h_k,a^*_k\}))|x_{k+1}=\bar{b}]  &&\\
& + \left(1-\frac{\Bar{P}(\bar{b},s_k)}{\sum_{b\in \mathcal{B}(s_k)\cap (a^*_k)^c} \Bar{P}(b,s_k)}\right) &&\\
&\hspace{1em}{\mathbb{E}[V^{\pi^*}((x_{l(k)},n(k)+1,\{h_k,a^*_k\}))|x_{k+1} \in \mathcal{B}(s_k)\setminus\{a^*_k \cup \bar{b}\}]}
\end{aligned}
\end{equation}
By Lemma~\ref{lem:tower_bound}, ~\eqref{eq:tower_equation} can be upper bounded by, 
\begin{equation*}
\begin{aligned}
&V^{\pi^*}((x_{l(k)},n(k)+1,\{h_k,a^*_k\})) &&\\
 &\leq \frac{\Bar{P}(\bar{b},s_k)}{\sum_{b\in \mathcal{B}(s_k)\cap (a^*_k)^c} \Bar{P}(b,s_k)}\mathbb{E}[V^{\pi^*}((\bar{b},0,\emptyset))|x_{k+1}=\bar{b}]  &&\\
&  + \left(1-\frac{\Bar{P}(\bar{b},s_k)}{\sum_{b\in \mathcal{B}(s_k)\cap (a^*_k)^c} \Bar{P}(b,s_k)}\right)&&\\
&{\mathbb{E}[V^{\pi^*}((x_{l(k)},n(k)+1,\{h_k,a^*_k\cup \bar{b}\}))|x_{k+1} \in \mathcal{B}(s_k)\setminus\{a^*_k \cup \bar{b}\}]}
\end{aligned}
\end{equation*}
Thus, we upper bound~\eqref{eq:m1} as,
\begin{equation*}
\begin{aligned}
&V^{\pi^*}(s_k)  < \bar{P}(\bar{b},s_k)r-c+\sum_{b^j\in a^*_k}(\bar{P}(b^j,s_k)r-c) &&\\
 &+ \gamma \sum_{b^j \in a^*_k}\bar{P}(b^j,s_k)V^{\pi^*}((b^j,0,\emptyset))
+\gamma \left( \sum_{b^j \notin a^*_k}\bar{P}(b^j,s_k)\right) &&\\
 &\Bigg(\frac{\Bar{P}(\bar{b},s_k)}{\sum_{b\in \mathcal{B}(s_k)\cap (a^*_k)^c} \Bar{P}(b,s_k)}\mathbb{E}[V^{\pi^*}((\bar{b},0,\emptyset))|x_{k+1}=\bar{b}]  &&\\
 &+ \left(1-\frac{\Bar{P}(\bar{b},s_k)}{\sum_{b\in \mathcal{B}(s_k)\cap (a^*_k)^c} \Bar{P}(b,s_k)}\right)&&\\
 &{\mathbb{E}[V^{\pi^*}((x_{l(k)},n(k)+1,\{h_k,a^*_k\cup \bar{b}\}))|x_{k+1} \in \mathcal{B}(s_k)\setminus\{a^*_k \cup \bar{b}\}]} \Bigg). &&\\
\end{aligned}
\end{equation*}
Since $ \sum_{b^j \notin a^*_k}\bar{P}(b^j,s_k) ={\sum_{b\in \mathcal{B}(s_k)\cap (a^*_k)^c} \Bar{P}(b,s_k)}$,
%
%
\begin{equation*}
\begin{aligned}
V&^{\pi^*}(s_k) <\bar{P}(\bar{b},s_k)r-c+\sum_{b^j\in a^*_k}(\bar{P}(b^j,s_k)r-c) &&\\
&+\gamma \sum_{b^j \in a^*_k}\bar{P}(b^j,s_k)V^{\pi^*}((b^j,0,\emptyset))
&&\\
& +\gamma \left( \Bar{P}(b,s_k)\mathbb{E}[V^{\pi^*}((\bar{b},0,\emptyset))|x_{k+1}=\bar{b}]\right) &&\\
& + \gamma\left(\sum_{b^j \notin a^*_k}\bar{P}(b^j,s_k)-\Bar{P}(\bar{b},s_k)\right)&&\\
&{\times \mathbb{E}[V^{\pi^*}((x_{l(k)},n(k)+1,\{h_k,a^*_k\cup \bar{b}\}))|x_{k+1} \in \mathcal{B}(s_k)\setminus\{a^*_k \cup \bar{b}\}]}. 
\end{aligned}
\end{equation*}
Hence,
\begin{equation*}
\begin{aligned}
 &V^{\pi^*}(s_k) < \bar{P}(\bar{b},s_k)r-c+\sum_{b^j\in a^*_k}(\bar{P}(b^j,s_k)r-c) &&\\
 &\nonumber+\gamma \sum_{b^j \in a^*_k}\bar{P}(b^j,s_k)V^{\pi^*}((b^j,0,\emptyset))
&&\\
& \nonumber+\gamma  \Bar{P}(\bar{b},s_k)V^{\pi^*}((\bar{b},0,\emptyset)) + \gamma\left(\sum_{b^j \notin a^*_k\cup \bar{b}}\bar{P}(b^j,s_k)\right) &&\\
&\nonumber\hspace{3em}\times V^{\pi^*}((x_{l(k)},n(k)+1,\{h_k,a^*_k\cup \bar{b}\})). &&\\
& \nonumber =\sum_{b^j\in a^*_k\cup \bar{b}}\big((\bar{P}(b^j,s_k)r-c)+ \gamma \bar{P}(b^j,s_k)V^{\pi^*}((b^j,0,\emptyset))\big)
&&\\
&\nonumber{ + \gamma\left(\sum_{b^j \notin a^*_k\cup \bar{b}}\bar{P}(b^j,s_k)\right)V^{\pi^*}((x_{l(k)},n(k)+1,\{h_k,a^*_k\cup \bar{b}\}))}&&\\
&\nonumber =Q^{\pi^*}(s_k,a^*_k\cup \bar{b}) 
\end{aligned}
\end{equation*}
We define a policy $\Tilde{\pi}$ such that $\Tilde{\pi}(s_k) = a^*_k \cup \Bar{b}$. For all the other states, $\Tilde{\pi} = \pi^*$. Thus, by the definition of the value function, 
\begin{equation*}
\begin{aligned}
    V^{\pi^*}(s) &<  V^{\Tilde{\pi}}(s),\  s = s_k\quad \mathrm{and}\quad
  V^{\pi^*}(s) &\leq  V^{\Tilde{\pi}}(s), \  s \neq s_k. &&\\
  \end{aligned}
\end{equation*}
This violates the optimality of $\pi^*$. Hence all sensors with $\bar{P}(\bar{b},s_k) > \frac{c}{r}$ must be turned on by the optimal policy $\pi^*$.
\end{proof}
\subsection{The $\mathrm{Q}_{\mathrm{MDP}}$ policy} \label{appendix:qmdp}
\begin{lemma}\label{lem:qmdp_pol}
    The policy by the $\mathrm{Q}_{\mathrm{MDP}}$ algorithm  $\pi^{\mathrm{Q}_{\mathrm{MDP}}}$,  at time $k+1$, will turn on the sensors at all $b^j$ where $j$ satisfies~\eqref{eq:qmdp_policy}
    where $v_k= (\beta_k,n(k))$ is the belief at time $k$.   
\end{lemma}
\begin{proof}
The policy $\pi_{\mathrm{Q}_{\mathrm{MDP}}}$~\cite{littman_qmdp} assumes that the state is perfectly observable after a control action $a$ is taken and hence the $\mathrm{Q}_{\mathrm{MDP}}$ action when the belief is $v_k$ is the action that maximizes the equation 
\begin{equation}\label{eq:Q_MDP}
\begin{aligned}
\max_{a \in \mathcal{A}} Q(v_k,a) =\max_{a \in \mathcal{A}}&\sum_{j:b^j\in a}([\beta_k P]_j r-c)&&\\
&+\gamma \sum_{j}[\beta_k P]_j V^{*}_{\mathrm{Q}_{\mathrm{MDP}}}(\mathbf{e}_{b^j}),
\end{aligned}
\end{equation}
when $n(k) \leq T_{\max}$.
The action does not affect the second term, so if $v_k$ is the belief at time $k$, the maximizing action, which is a solution to~\eqref{eq:Q_MDP} turns on all sensors $b^j$ at time $k+1$ such that $\{b^j: [v_k P]_j \geq \frac{c}{r},\ j\in [N]^2]\}$. 
When $n(k) = T_{\max}+1$, the action is constrained to $\hat{a}_s$.
\end{proof}
\begin{remark}
Multiple actions may lead to the maximization of Equation~\eqref{eq:Q_MDP} at belief $v_k$. However, a proof analogous to Lemma~\ref{lem:tower_bound} can be established for $\mathrm{Q}_{\mathrm{MDP}}$, demonstrating that the value function of the policy~\eqref{eq:qmdp_policy} serves as an upper bound for all policies maximizing~\eqref{eq:Q_MDP}.
\end{remark}
\subsection{Upper Bound \texorpdfstring{$V^*_{\mathrm{Q}_{\mathrm{MDP}}}$}{TEXT}}~\label{sec:upper_bound}
Consider the definition
\begin{equation}~\label{eq:upper_bound_v}
\begin{aligned}
V^{*}_{\mathrm{Q}_{\mathrm{MDP}}}(v_k) \triangleq\max_{a \in \mathcal{A}}&\sum_{j:b^j\in a}([\beta_k P]_j r-c)&&\\
&+\gamma \sum_{j}[\beta_k P]_j V^{*}_{\mathrm{Q}_{\mathrm{MDP}}}(\mathbf{e}_{b^j}),
\end{aligned}
\end{equation}
$V^{*}_{\mathrm{Q}_{\mathrm{MDP}}}(v_k)$ is an upper bound for $V^{\pi^*_{\mathrm{POMDP}}}(v_k)$ since the state is assumed to be observable after taking the control action $a$. To solve for $V^{*}_{\mathrm{Q}_{\mathrm{MDP}}}(v_k)$, a system of equations for $V^*_{\mathrm{Q}_{\mathrm{MDP}}}(e_{b^j})$ is derived from \eqref{eq:upper_bound_v} and solved for  $V^*_{\mathrm{Q}_{\mathrm{MDP}}}(e_{b^j})$ for all $j \in [N^2]$. Subsequently $V^*_{\mathrm{Q}_{\mathrm{MDP}}}(e_{b^j})$ is substituted in \eqref{eq:upper_bound_v} to solve for $V^*_{\mathrm{Q}_{\mathrm{MDP}}}(v_k)$. This is further elaborated in~\cite{ross2008online}.
\subsection{Proof of Theorem~\ref{thm:estimator}} \label{proof:estimator}
\begin{proof}
\begin{align*}
Q^{\pi^*}(s_k, \tilde{a})
&=\sum_{b^j\in \tilde{a}}(\bar{P}(b^j,s_k)r-c) &&\\
&\hspace{1em}+\gamma \sum_{b^j \in \tilde{a}}\bar{P}(b^j,s_k)V^{\pi^*}((b^j,0,\emptyset))
\end{align*}
\begin{equation*}
\begin{aligned}
Q^{\pi^*}(s_k, \tilde{a}_x)
&=\sum_{b^j \in \tilde{a}\setminus x}(\bar{P}(b^j,s_k)r-c) &&\\
&+\gamma \sum_{b^j \in \tilde{a}\setminus x}\bar{P}(b^j,s_k)V^{\pi^*}((b^j,0,\emptyset)) &&\\
&+\gamma \bar{P}(x,s_k)V^{\pi^*}((x_{l(k)},n(k)+1,\{h_k,(\tilde{a}\setminus x)\}))
\end{aligned}
\end{equation*}
It can be easily shown that for the optimal policy $\pi^*$ on the Track-MDP, when the target is at $x$, 
\begin{equation}\label{eq:est_st_val}
V^{\pi^*}((x_{l(k)},n(k)+1,\{h_k,(\tilde{a}\setminus x)\})) = V^{\pi^*}((x,0,\emptyset)).  
\end{equation}
This is true since if all sensors except at $x$ are turned on by action $\tilde{a}_x$ and the target is not detected, the only position of the target is $x$, which makes~\eqref{eq:est_st_val} true. Thus,
\begin{equation*}
\begin{aligned}
 \bar{P}(b^j,s_k) =\frac{Q^{\pi^*}(s_k, \tilde{a})  - Q^{\pi^*}(s_k, \tilde{a}_x) + c}{r}. 
\end{aligned}
\end{equation*}
Since $\mathrm{Pr}(X_{k+1}=b^j|s_k ) = \bar{P}(b^j,s_k)$ and the sensors at positions $a_k$ were turned on at time $k+1$ and the target was unobserved, 
\begin{equation*}
    \hat{X}_{k+1} = \arg \max_{x \in \mathcal{X}, x\notin a_k} \frac{Q^{\pi^*}(s_k, \tilde{a})  - Q^{\pi^*}(s_k, \tilde{a}_x) + c}{r}
\end{equation*}
\end{proof}
\subsection{Proof of Theorem~\ref{thm:optimalaspomdp}} \label{proof:optimalaspomdp}
\begin{proof}
The POMDP has the following state space.
\begin{equation}\label{eq:pomdp_aug_state}
\begin{aligned}
   \tilde{\mathcal{S}} = \{(x,n)| x \in \mathcal{X} \ \text{and} \ 0\leq n \leq T_{\max}+1\}
\end{aligned}    
\end{equation}
The action space of the POMDP is the same set $\mathcal{A}_{\mathcal{X}}$ defined in Section~\ref{sec:tmdp_for} and the action is constrained to $\hat{a}_s$ if $n= T_{\max}+1$. The reward of the POMDP is the same as the pseudo-reward as defined in Equation~\ref{eq:reward_func}. The state transition of the POMDP is defined as follows. If the state is $\tilde{s}_k = (x_k,n)$ at time $k$, then at time $k+1$ the state evolves as
\begin{equation}\label{eq:pom_trans_def}
 \tilde{s}_{k+1} = \begin{cases}
     (x_{k+1},n+1)& \text{if , $x_{k+1} \notin a_k$ and $n \leq T_{\max} $}\\
     (x_{k+1},0)& \text{if $\tilde{s}_{k} = (x_k,T_{\max}+1)$ }, \\
     (x_{k+1},0)& \text{if $x_{k+1} \in a_k$ }, \\
     (b^T,0)& \text{if $x_{k+1} = b^T$ }, \\
 \end{cases}  , 
\end{equation}
The transition probability $P(\tilde{s}_{k+1}|\tilde{s}_{k},a_k)$ is induced by the target transition probability matrix $P$ defined in Equation~\ref{eq:process_st_transition}.

The observation model of the POMDP on the state space $\tilde{\mathcal{S}}$ is $y_{k+1} = x_{k+1} $ if $\tilde{s}_{k+1} = (x_{k+1},0)$ and $y_{k+1} = b^T $ if $x_{k+1} = b^T$ and $ y_{k+1} = \mathcal{E}$ if otherwise.
For the POMDP defined by~\eqref{eq:pomdp_aug_state},\eqref{eq:pom_trans_def} and the defined observation model, all policies on the POMDP are $T_{\max}$ safe since if $\tilde{s}_{k} = (x_k, T_{\max} +1)$, then $\tilde{s}_{k+1} = (x_{k+1}, 0)$ and thus $y_{k+1} = x_{k+1}$. 

If a POMDP policy is constrained to take the action $\hat{a}_s$ if unobserved for $T_{\max}+1$ time steps, the sufficient statistic at time $k$ is $v_k= (\beta_k,n(k))$. Hence the POMDP policy must have a belief over the state space $\Tilde{\mathcal{S}}$. The optimal POMDP policy among the set of POMDP policies that are constrained to take the action $\hat{a}_s$ is $\pi^*_{\mathrm{POMDP}}$.
By the definition of the sensing grid in Section~\ref{sec:sensing_grid} and its initialization, once the target enters the grid, its initial belief is an element of $\mathcal{B}_{\mathrm{Initial}}$ defined as 
\begin{align*}
    \mathcal{B}_{\mathrm{Initial}} \triangleq \{(e_{b^j},0) | b^j \in \mathcal{X} \},
\end{align*}
The belief $(e_{b^j},0)$ for the POMDP essentially means the target is at $b^j$ at time $k=0$. This essentially maps to the state $s_0 = (b^j,0,\emptyset)$ of the Track-MDP.
Define an episode of a policy $\pi$ on the POMDP with initial belief $v_0$ as 
\begin{align*}
    \{v_0,\pi(v_0),y_1,v_1,\pi(v_0), y_2,\cdots \}.
\end{align*}
Hence an episode of the POMDP with initial belief $v_0 \in \mathcal{B}_{\mathrm{Initial}}$ and policy $\pi^*_{\mathrm{POMDP}}$ is given by~\eqref{eq:epi_pom}. Since $v_0$ is mapped to $s_0$, the equivalent episode on the Track-MDP is~\eqref{eq:epi_track}.
\begin{align}\label{eq:epi_pom}
\{v_0,\pi^*_{\mathrm{POMDP}}(v_0),y_1,v_1,\pi^*_{\mathrm{POMDP}}(v_1),y_2,\cdots \}. &&\\
\{s_0,\pi^*_{\mathrm{POMDP}}(v_0),y_1,s_1,\pi^*_{\mathrm{POMDP}}(v_1),y_2,\cdots \}~\label{eq:epi_track}
\end{align}
The rewards for the episodes~\eqref{eq:epi_pom},~\eqref{eq:epi_track} are the same since initial beliefs are the same and the actions and the observations are the same. Hence,
\begin{align*}
V_{\mathrm{POMDP}}^{\pi^*_{\mathrm{POMDP}}} = V_{\mathrm{TMDP}}^{\pi^*_{\mathrm{POMDP}}}
\end{align*}
If $\pi^{*}_{\mathrm{TMDP}}$ is the optimal policy for the TrackMDP, 
$V_{\mathrm{TMDP}}^{\pi^{*}_{\mathrm{POMDP}}} \leq V_{\mathrm{TMDP}}^{\pi^{*}_{\mathrm{TMDP}}} $.
Since the Track-MDP uses just the observations $y_k$, it is a method to solve the POMDP and hence
\begin{align*}
V_{\mathrm{TMDP}}^{\pi^{*}_{\mathrm{TMDP}}} = V_{\mathrm{POMDP}}^{\pi^*_{\mathrm{POMDP}}}.
\end{align*}
\end{proof}

\end{document}